\documentclass[twocolumn]{aastex63}

\usepackage{amsmath}
\newcommand{\bs}{\boldsymbol}
\newcommand{\rd}{{\rm d}}

\received{June 1, 2019}
\revised{January 10, 2019}
\accepted{\today}
\submitjournal{ApJ}

\shorttitle{Deep Learning of the Eddington Tensor}
\shortauthors{Harada et al.}
\graphicspath{{./}{figures/}}

\begin{document}

\title{Deep Learning of the Eddington Tensor in the Core-collapse Supernova Simulation}

\correspondingauthor{Akira Harada}
\email{akira.harada@riken.jp}

\author[0000-0003-1409-0695]{Akira Harada}
\affiliation{Interdisciplinary Theoretical and Mathematical Sciences Program (iTHEMS), RIKEN, Wako, Saitama 351-0198, Japan}

\author{Shota Nishikawa}
\affiliation{Advanced Research Institute for Science and Engineering, Waseda University, 3-4-1 Okubo, Shinjuku, Tokyo 169-8555, Japan}


\author{Shoichi Yamada}
\affiliation{Advanced Research Institute for Science and Engineering, Waseda University, 3-4-1 Okubo, Shinjuku, Tokyo 169-8555, Japan}



\begin{abstract}
We trained deep neural networks (DNNs) as a function of the neutrino energy density, flux, and the fluid velocity to reproduce the Eddington tensor for neutrinos obtained in our first-principles core-collapse supernova (CCSN) simulations. Although the moment method, which is one of the most popular approximations for neutrino transport, requires a closure relation, none of the analytical closure relations commonly employed in the literature captures all aspects of the neutrino angular distribution in momentum space. In this paper, we developed a closure relation by using the DNN that takes the neutrino energy density, flux, and the fluid velocity as the input and the Eddington tensor as the output. We consider two kinds of DNNs: a conventional DNN named a component-wise neural network (CWNN) and a tensor-basis neural network (TBNN). We found that the diagonal component of the Eddington tensor is reproduced better by the DNNs than the M1-closure relation especially for low to intermediate energies. For the off-diagonal component, the DNNs agree better with the Boltzmann solver than the M1 closure at large radii. In the comparison between the two DNNs, the TBNN has slightly better performance than the CWNN. With the new closure relations at hand based on the DNNs that well reproduce the Eddington tensor with much smaller costs, we opened up a new possibility for the moment method.

\end{abstract}



\section{Introduction} \label{sec:intro}
The core-collapse supernova (CCSN) is the explosive death of a massive star \citep{1934PNAS...20..254B}.
The explosion energy is $\sim 10^{51}\,{\rm erg}$ and the ultimate energy source is the gravitational energy released when a stellar core collapses to form a neutron star.
The recent discovery of the binary neutron star merger revealed that it is the site of the $r$-process nucleosynthesis \citep{2017PASJ...69..102T}. To understand the chemical evolution of the universe in a coherent way, the understanding of the neutron star formation event, i.e., the CCSN is important.

The leading hypothesis of the explosion mechanism of the CCSN is the neutrino heating mechanism \citep[see, e.g.,][for a review]{2012ARNPS..62..407J}. This scenario is illustrated as follows. At the end of the stellar evolution, the stellar core experiences gravitational collapse. When the central density reaches the nuclear density, the inner part of the core becomes stiff owing to nuclear repulsive forces and bounces. Then, the bounce shock is formed at the interface between the subsonic inner and supersonic outer cores. This bounce shock loses its energy as it propagates and stalls eventually. After the core bounce, a proto-neutron star (PNS) is formed at the center. Energetic neutrinos are copiously emitted from the PNS. A part of these neutrinos are absorbed by matter behind the shock and energizes it. At last, the shock is supposed to revive.

Various physical processes are involved in the neutrino heating mechanism. To investigate this mechanism, numerical simulations are utilized. The spherically symmetric simulations were conducted first and concluded that they do not give a successful shock revival except for the lightest progenitor \citep{2001PhRvD..63j3004L, 2005ApJ...629..922S, 2006A&A...450..345K, 2020arXiv201016254M}. Then, multidimensional simulations have been performed \citep[e.g.,][]{2020MNRAS.491.2715B, 2021ApJ...915...28B} and produced shock revival indeed in many cases thanks to the help of multidimensional effects such as turbulence.

The conclusions may be still revised. In fact, although shock revivals are observed in these multi-dimensional simulations, it takes them a long time to reach the explosion energy as observed and it is unclear if they are able to synthesize a sufficient amount of ${}^{56}{\rm Ni}$ \citep{2019MNRAS.483.3607S, 2019ApJ...886...47S, 2021ApJ...908....6S}. Besides, the neutrino transport is approximated one way or another in most of the simulations to reduce the numerical cost \citep{2006ApJ...640..878B, 2012ApJ...755...11K, 2002A&A...396..361R, 2009ApJ...698.1174L}. Different approximation methods are supposed to contribute to differences in the simulation outcomes (e.g., whether the shock revives or not).

To change the situation, we have developed and run a Boltzmann-radiation-hydrodynamics code, which simultaneously solves the hydrodynamics equations and the Boltzmann equation for neutrinos. No artificial approximation except for the mandatory discretization is employed in this code. So far, several CCSN simulations using this code have been reported \citep{2018ApJ...854..136N, 2019ApJ...880L..28N, 2019ApJ...872..181H, 2020ApJ...902..150H, 2020ApJ...903...82I}.

\citet{2019ApJ...872..181H, 2020ApJ...902..150H} calculated the second angular moment of the distribution function of neutrinos, or the Eddington tensor, and compared it with the M1-closure prescription, one of the most popular closure relations at present. They found some discrepancies between the two results possibly due to a kind of the ray-collision\footnote{This is one of the well-known drawbacks of the M1-closure prescription. When two rays cross in vacuum, they should go through if there is no interaction. In the M1-closure prescription, however, the rays collide with each other artificially, and the subsequent propagation becomes unphysical.}. This suggests that the information on the fluid, e.g., the matter velocity distribution, might be useful to improve the M1-closure scheme. Besides, the contribution of the fluid velocity to the Eddington tensor in the closure relation seems to be overestimated in the optically thick regime and need more careful treatment. One way to improve this situation is to develop a closure relation calibrated by the Boltzmann-radiation-hydrodynamics simulations.

In this paper, we utilize the machine learning technique with the deep neural network \citep[DNN; see][for a review]{2015Natur.521..436L} to obtain the simulation-calibrated closure relation. Since this is the first-ever attempt of the sort, the main goal of this paper is to give a proof of principle, i.e., to demonstrate that the neural networks can be trained indeed so that it could give an estimate of the Eddington tensor from the neutrino energy density, flux, and the fluid velocity; it could include other quantities, which is beyond the scope of this paper, though. The machine learning with the DNN is a powerful tool to estimate such a complicated functional relation. Thanks to the universal approximation theorem \citep{Cybenko:1989aa}, any function can be expressed by a neural network with infinite number of nodes. It is well known that the DNN can approximate complicated functions fairly well even with a finite number of nodes. If there is indeed some functional relation between the Eddington tensor and other quantities, we can hence expect that the DNN will provide us with a reasonable approximation.

The deep learning technique has a wide variety of applications. One example is the image recognition. As for the astrophysical applications, classifications of the types of supernovae and other optical transients are suggested \citep{2016ApJS..225...31L, 2020PASJ..tmp..240T}. \citet{ling_kurzawski_templeton_2016} proposed, on the other hand, an application of the deep learning technique to a closure relation for the Reynolds-decomposed quantities in hydrodynamical simulations of turbulence. We make a similar attempt for the closure relation for neutrino transport in the CCSN simulation in this paper. Recently, some similar attempts to give closure relations for moment-scheme radiation/particle transport were proposed \citep{2021arXiv210505690H, 2021arXiv210514410H, 2021arXiv210900700H, 2021arXiv210608973P, 2021arXiv210609445S}, respecting the hyperbolicity condition to obtain stable schemes. However, they either sacrificed the conservation law of the moments or assumed the maximum-entropy condition, which does not hold in CCSN simulations \citep{2021arXiv210905846I}. Therefore, other approaches are required for neutrino transport in the CCSN simulations.

This paper is organized as follows. In section \ref{sec:method}, we review the DNN briefly and present the network we employ in this work. The data used in this paper are also explained there. Then, we present the results in section \ref{sec:results}. Finally, in section \ref{sec:concl}, we conclude this paper with a summary of what we have done. The metric signature is $(-+++)$.

\section{Deep Neural Network} \label{sec:method}

\subsection{Supervised Deep Learning}
The DNN is a multi-input and multi-output composite function composed of affine transformations and nonlinear activation functions. The basic unit of the DNN is an artificial neuron, which is a multi-input single-output function, and called a node. The functional form of the artificial neuron is $z = \sigma({\bs w} \cdot  {\bs x}+ b)$, where $\bs x$, $z$, $\bs w$, and $ b$ are the input, output, weight, and bias, respectively. The bold symbols are vectors.\footnote{This is just in mathematical sense: they are not necessarily related to the spacetime dimension.} The function $\sigma$ is non-linear and called the activation function. A layer is defined as a vectorized artificial neurons: a multi-input and multi-output function of the form of $z_i = \sigma(\sum_j W_{ij} x_j + b_i)$, where $W$ is the weight matrix, and the subscripts $i$, $j$ indicate the component of the vectors and the matrix. Finally, the DNN is defined as the composition of the layers: $z_i = \sigma(\sum_j W^L_{ij} \sigma(\sum_k W^{L-1}_{jk} \sigma(\cdots \sum_n W^2_{\ell n}\sigma(\sum_m W^1_{nm}x_m+b^1_n) + \cdots) + b^{L-1}_j) + b^L_i)$. The superscripts represent the ID numbers of the layers.

The supervised learning, which is employed in this paper, is to tune the weights and biases of the DNN to reproduce the relation of the input and the output of a given data set. 
Let us consider a set of input--output pairs $\{({\bs x}^{\rm sv}_{(a)},{\bs z}^{\rm sv}_{(a)})\}$, where the subscripts indicate the ID numbers of the pairs. The prediction ${\bs z}^{\rm pd}_{(a)}$ is the output of the DNN when $\bs x^{\rm sv}_{(a)}$ is the input. The loss function $\mathcal{L}$ measures how close the prediction is to the true output and is defined as $\mathcal{L} := \sum_{a} ({\bs z}^{\rm pd}_{(a)} - {\bs z}^{\rm sv}_{(a)})^2/\sum_{a}1$, i.e., the mean square of the difference between the two, where $a$ runs over the data-pair ID. The weights and biases are iteratively tuned to decrease $\mathcal{L}$ as follows: $W_{ij}^{(t+1)} = W_{ij}^{(t)} - \eta (\partial \mathcal{L}/\partial W_{ij})^{(t)}$, where $\eta$ and the superscripts are the learning rate and the ID numbers of the iteration, respectively. The bias $b$ is similarly updated. 
The data are divided into minibatches with sizes $N_{\rm B}$. For each step of the iteration, $\mathcal{L}$ is evaluated for all data in one minibatch (the data ID $a$ runs from $1$ to $N_{\rm B}$) and the parameters are updated. Then, another step with a different minibatch follows. After all the minibatches, and hence all the data, are used to update the parameters, this procedure is repeated over from the first minibatch; this one cycle is called the epoch. Because the evaluation of $\mathcal{L}$ is based on the data sampling, this method is called the stochastic gradient descent (SGD) method. 

The back-propagation is an efficient technique to calculate the gradient $\partial \mathcal{L}/\partial W^\ell_{ij}$. We define $u^\ell_{j}$ and $z^\ell_{j}$ as the augment of the activation function and the output at the $\ell$-th layer of the DNN, respectively: $u^\ell_{j} := \sum_k W^\ell_{jk}\sigma(\sum_i W^{\ell-1}_{ki}\sigma(\cdots \sigma(\sum_m W^1_{nm}x_m + b_n^1)+\cdots)+b^{\ell-1}_k)+b^\ell_j$; $z^\ell_{j} := \sigma(u^\ell_{j})$. The gradient of the loss function is then expressed as
\begin{equation}
\frac{\partial \mathcal{L}}{\partial W^\ell_{ij}} = \frac{\partial \mathcal{L}}{\partial u^\ell_j}\frac{\partial u^\ell_j}{\partial W^\ell_{ij}} = \delta^\ell_j z^{\ell-1}_i, \label{eq:gradij}
\end{equation}
since $u^\ell_j = \sum_i W^\ell_{ji}z^{\ell-1}_i$, where $\delta^\ell_j := \partial \mathcal{L}/\partial u^\ell_j$ is called the delta. Note that we do not take the sum with respect to the repeated index $\ell$ in equation (\ref{eq:gradij}). This delta is calculated sequentially backward from the output layer $\ell=L$ to the input layer $\ell=1$ as
\begin{equation}
\delta^\ell_j = \sum_k \delta^{\ell+1}_k W^{\ell+1}_{kj} \sigma^\prime(u_j^\ell).
\end{equation}
The primed symbol stands for the derivative of the function. With this backwardly propagating delta, we obtain the gradient $\partial \mathcal{L}/\partial W^\ell_{ij}$ from equation (\ref{eq:gradij}).

The batch normalization technique \citep{2015arXiv150203167I} is one of the powerful tools for learning. The so-called batch normalization layer is utilized to suppress possible influences of the internal covariate shift, which is the change in the statistical properties of the layer output among different update steps. In this layer, the input data are averaged over the minibatch, and they are shifted and normalized to have the same average and variance over the different minibatches. The batch normalization sometimes plays a key role in the successful learning.

\subsection{Data Description}
The input--output data pairs employed in this paper are taken from a 2D axisymmetric CCSN simulation with the Boltzmann-radiation-hydrodynamics code. This code solves the Boltzmann equations for neutrino transport, the Newtonian hydrodynamics equations, and the Poisson equation for Newtonian gravitational potential simultaneously. The detailed descriptions of the code are presented in \citet{2012ApJS..199...17S, 2014ApJS..214...16N, 2017ApJS..229...42N, 2019ApJ...878..160N}. The nuclear equation of state (EOS) employed is the Furusawa--Togashi EOS based on the variational method at supra-nuclear densities \citep{2017NuPhA.961...78T} and the nuclear statistical equilibrium description at sub-nuclear densities \citep{2017JPhG...44i4001F}. The neutrino interactions are the same as those in \cite{2019ApJ...880L..28N}. The progenitor is the nonrotating $15\,M_\odot$ model taken from \citet{2002RvMP...74.1015W}. Although three kinds of neutrinos (the electron-type neutrinos, its antineutrinos, and the heavy-lepton-type neutrinos) are considered in the simulation, we focus only on the electron-type neutrinos in this paper. The simulation domain covers the full meridian plane up to $5000\,{\rm km}$, and the neutrino energy is considered up to $300\,{\rm MeV}$. The numbers of the radial, zenith, energy, and momentum-angle grid points are $(N_r,N_\theta, N_\epsilon, N_{\theta_\nu}, N_{\phi_\nu})=(384,128,20,10,6)$. The dynamical features and the neutrino distributions will be reported elsewhere. 
The simulation results on each grid point in space and energy are fed to the DNN with the fluid velocity, the neutrino energy density and flux as input and the Eddington tensor as output.

The Eddington tensor is the second angular moment of the distribution function of neutrinos divided by the zero-th moment. In this paper, we follow the definition in \citet{2019ApJ...872..181H} based on \citet{1981MNRAS.194..439T, 2011PThPh.125.1255S}. In the remainder of this subsection, we employ the units with $c=1$ and the Greek and Latin indices running over $0$--$3$ (spacetime) and $1$--$3$ (space), respectively. First, we define the unprojected second moment of the distribution function as
\begin{eqnarray}
M^{\alpha\beta}(\epsilon) &:=& \int f \delta\left(\frac{\epsilon^3}{3} - \frac{\epsilon^{\prime 3}}{3}\right) p^{\prime \alpha} p^{\prime \beta} {\rm d}V_{p^\prime},
\end{eqnarray}
where $f$, $p^\prime$, $\rd V_{p^\prime}$, $\epsilon^\prime = -p^\prime \cdot u$, and $u$ are the distribution function and 4-momentum of neutrinos, the invariant volume element in the momentum space, the neutrino energy measured in the fluid-rest frame, and the 4-velocity of matter, respectively.
From $M^{\alpha\beta}$, we obtain the second and zero-th angular moments by spatial--spatial and temporal--temporal projections, respectively, as formulated later.

We evaluate the Eddington tensor in four different ways: (1) $k_{\rm Boltz,FR}^{ij}$ is calculated from $M^{\alpha\beta}$ in the fluid-rest frame, (2) $k_{\rm Boltz,LB}$ is obtained from $M^{\alpha\beta}$ in the same way but in the laboratory frame, (3) $k_{\rm M1,FR}^{ij}$ is an approximation based on the M1-closure prescription applied in the fluid-rest frame, and (4) $k_{\rm M1,LB}^{ij}$ is the same as in (3) but for the laboratory frame. We refer to $k_{\rm Boltz,FR/LB}^{ij}$ and $k_{\rm M1,FR/LB}^{ij}$ as the Boltzmann- and M1- Eddington tensors, respectively. The difference between $k_{\rm Boltz,FR}^{ij}$ and $k_{\rm Boltz,LB}^{ij}$ lies in the projection tensors: for the former the spatial projection tensor $h_\alpha{}^i = \delta_\alpha{}^i + u_\alpha u^i$ is employed with the temporal vector $u^\alpha$ to give $k_{\rm Boltz,FR}^{ij} := M^{\alpha\beta}h_\alpha{}^i h_\beta{}^j/E_{\rm FR}$, where $E_{\rm FR} = M^{\mu\nu}u_\mu u_\nu$ is the energy density in the fluid-rest frame; for the latter, on the other hand, the spatial projection tensor is $\gamma_\alpha{}^i = \delta_\alpha{}^i + n_\alpha n^i$ with the temporal vector $n^\alpha$ and the expression for the Eddington tensor is $k_{\rm Boltz,LB}^{ij} := M^{\alpha\beta}\gamma_\alpha{}^i \gamma_\beta{}^j/E_{\rm LB}$, where $E_{\rm LB} = M^{\mu\nu}n_\mu n_\nu$ is the energy density in the laboratory frame. Here, $\delta_\alpha{}^i$ is Kronecker's delta\footnote{The same symbol $\delta$ is used here and for the delta in the back-propagation. No confusion is expected, however, because the latter appears only in the explanation of the back-propagation in this paper.}, and $n^\alpha$ is a unit vector normal to the hypersurface with the constant time. 

The M1-Eddington tensor is constructed by the interpolation of the two limiting cases, i.e., the optically thick and thin limits. The second moment $P_{\rm M1,FR/LB}^{ij}$ is expressed first as
\begin{equation}
P_{\rm M1,FR/LB}^{ij} = \alpha P_{\rm thick,FR/LB}^{ij} + \beta P_{\rm thin,FR/LB}^{ij}, \label{eq:PM1}
\end{equation}
where $P_{\rm thick,FR/LB}^{ij}$ and $P_{\rm thin,FR/LB}^{ij}$ are the optically thick and thin limits of the Eddington tensor either in the fluid-rest or laboratory frames.
The coefficients $\alpha$ and $\beta$ are functions of the Eddington factor which is in turn given as a function of the flux factor $\tilde f$. The flux factor is defined as $\tilde f = |F_{\rm FR}|/E_{\rm FR}$ where $F_{\rm FR}^i = -M^{\mu\nu}h_\mu{}^i u_\nu$ is the energy flux in the fluid-rest frame. Note that the flux factor is always measured in the fluid-rest frame. The Eddington factor $\chi$ is originally defined as the eigenvalue of the Eddington tensor whose corresponding eigenvector is close to the flux. In the M1 prescription, however, it is given as a function of the flux factor \citep{1984JQSRT..31..149L}:
\begin{equation}
\chi = \frac{3+4\tilde{f}^2}{5+2\sqrt{4-3\tilde{f}}}. \label{eq:chi}
\end{equation}
With this Eddington factor, the coefficients in equation (\ref{eq:PM1}) are given as
\begin{eqnarray}
\alpha &=& \frac{3(1-\chi)}{2}, \label{eq:alpha} \\
\beta &=& \frac{3\chi -1}{2}. \label{eq:beta}
\end{eqnarray}
With the second moment $P_{\rm M1,FR/LB}^{ij}$ now at hand, the M1-Eddington tensors are defined as $k_{\rm M1,FR}^{ij} = P_{\rm M1,FR}^{ij}/E_{\rm FR}$ and $k_{\rm M1,LB}^{ij} = P_{\rm M1,LB}^{ij}/E_{\rm LB}$.

The limits of the second moment are given in the fluid-rest frame as follows:
\begin{eqnarray}
P_{\rm thick,FR}^{ij} &=& E_{\rm FR} \frac{\gamma^{ij}}{3}, \label{eq:PM1FRthick}\\
P_{\rm thin,FR}^{ij} &=& E_{\rm FR} \frac{F_{\rm FR}^i F_{\rm FR}^j}{|F_{\rm FR}|^2}. \label{eq:PM1FRthin}
\end{eqnarray}
In the laboratory frame on the other hand, we follow \citet{2011PThPh.125.1255S} and express them as
\begin{eqnarray}
P_{\rm thick,LB}^{ij} &=& E_{\rm FR} \frac{\gamma^{ij}+ 4 V^i V^j}{3} + F_{\rm FR}^i V^j + V^i F_{\rm FR}^j, \label{eq:PM1LBthick}\\
P_{\rm thin,LB}^{ij} &=& E_{\rm LB} \frac{F_{\rm LB}^i F_{\rm LB}^j}{|F_{\rm LB}|^2}, \label{eq:PM1LBthin}
\end{eqnarray}
where $F_{\rm LB}^i = -M^{\mu\nu}\gamma_\mu{}^i n_\nu$ and $V^i = u^i/u^0$ are the energy flux and
3-velocity in the laboratory frame, respectively.

These Eddington tensors are calculated for the result of the original simulation at $100\,{\rm ms}$ after bounce. Although the simulation domain extends to $r=5000\,{\rm km}$, we focus on the region $r\le 100\,{\rm km}$. Note that this radius is smaller than the shock radius at this time, $r_{\rm shock} \simeq 170\,{\rm km}$. The numbers of the radial, zenith, and energy grid points considered here are $\sim 200$, $\sim 130$, and $20$, respectively. Then, we have $\sim 540000$ data points. We randomly split this data into the training and validation sets. The training set constitutes $80\%$ of all the data and is used to train the DNN; the training set is further divided into minibatches with $N_{\rm B}=512$. The validation set is used to measure the performance of the DNN. Usually, the validation loss, i.e., the loss function calculated for the validation set is used to judge the performance. If the validation loss is small, the trained network well reproduces the functional relation between the input and the output even with the data that are not used to train the network. In this paper, however, we compare the output of the trained network applied for the validation set with the values of the Eddington tensor either obtained directly from the simulation or derived with the M1 prescription to the same input data.

\subsection{Structures of Deep Neural Network}
We consider two types of DNN in this paper: one is a simple DNN, which we call a Component-Wise Neural Network (CWNN) hereafter; the other is a Tensor-Basis Neural Network (TBNN) \citep{ling_kurzawski_templeton_2016}. We employ tensorflow \citep{tensorflow2015-whitepaper} with keras \citep{chollet2015keras} to implement and train these networks.

\begin{figure}[t]
\plotone{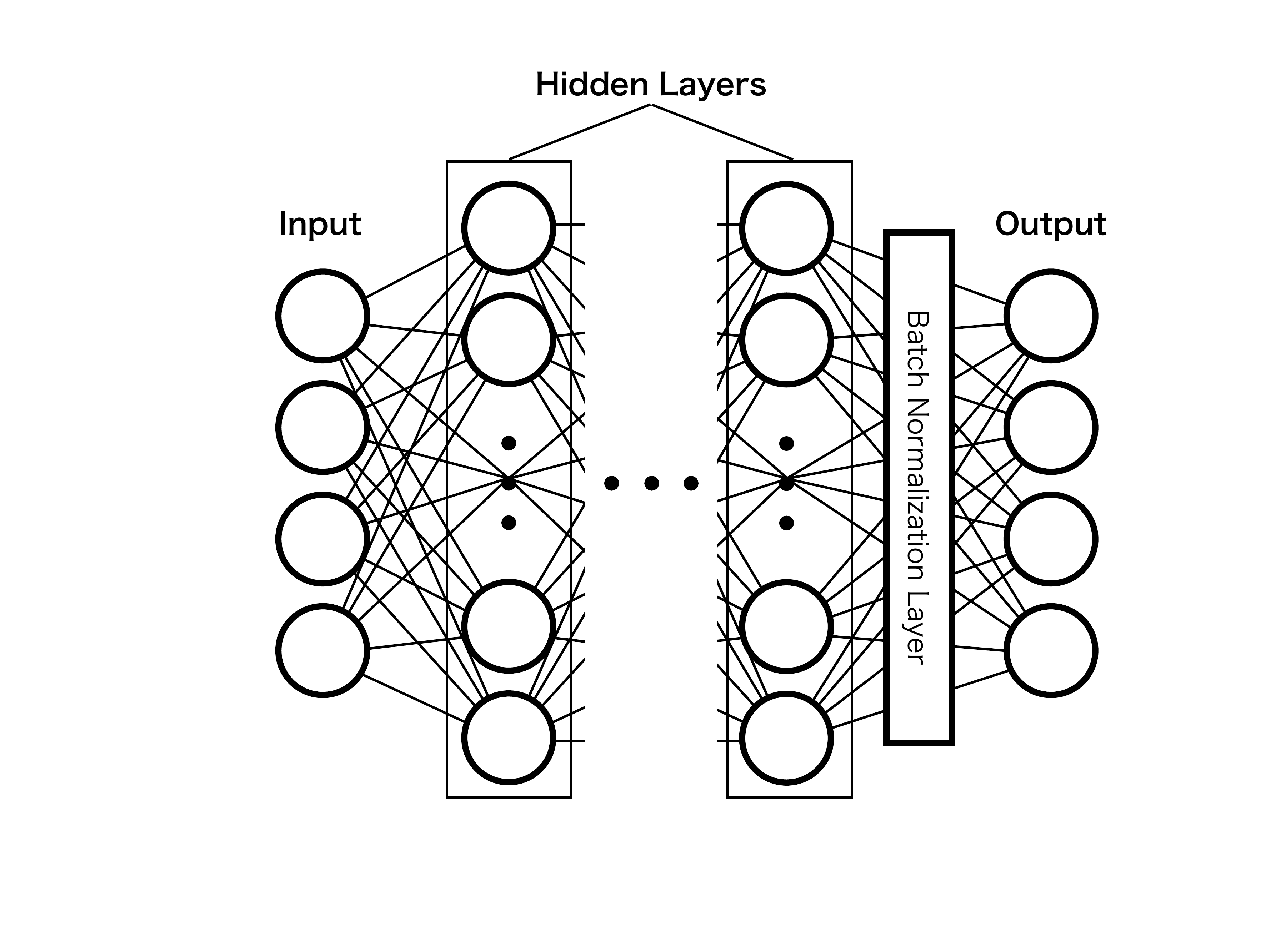}
\caption{The network diagram for the CWNN. Circles and lines connecting them represent the nodes and weights, respectively. The vertically aligned nodes form a layer. Data are processed from left to right, and hence the left- and right-end layers are the input and output layers, respectively. Other layers than the input and output are called the hidden layers as indicated by the thin rectangles. The batch normalization layer is inserted between the last-hidden and output layers as shown by the thick rectangle. \label{fig:CWNN}}
\end{figure}

The CWNN is illustrated in figure \ref{fig:CWNN} as a diagram. The node is represented by a circle. The lines connecting these circles express the weights. The data flows from left to right; the lines entering the circle from the left are the input to the node, wheres those emanating to the right are the output. Vertically arranged nodes compose a layer. Note that the layers other than the input and output layers are called the hidden layers. The left-most layer is the input layer which just outputs the input data. The right-most layer is the output layer which corresponds to $\ell = L$. The CWNN has a simple structure and outputs each component of the Eddington tensor directly individually: $k^{rr},k^{\theta\theta},k^{\phi\phi},k^{r\theta},k^{r\phi},k^{\theta\phi}$. The batch normalization layer is inserted between the final hidden layer and the output layer.

\begin{figure}[t]
\plotone{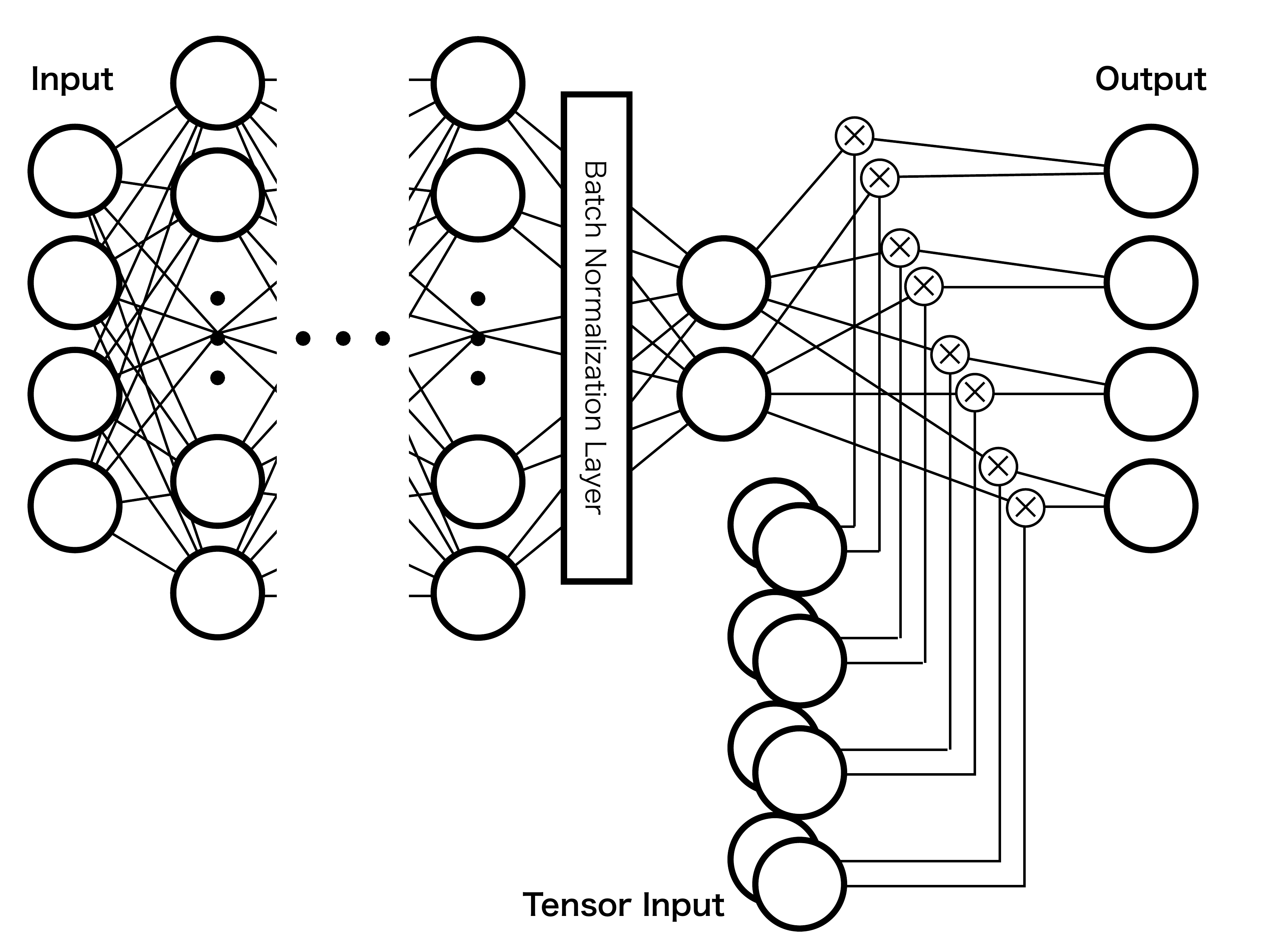}
\caption{The network diagram for the TBNN. The notation of the symbols (the circles, lines, and thick rectangle) is the same as in figure \ref{fig:CWNN}. Each node of the tensor input and output layers correspond to each component of the tensor. The symbol $\otimes$ represents the multiplication of the outputs. \label{fig:TBNN}}
\end{figure}

The structure of the TBNN is shown in figure \ref{fig:TBNN}. The TBNN has two kinds of input layer: the usual input layer and the tensor input layers. The tensor input layers have the same vector dimension as the output layer. Each node of the tensor input layer outputs each component of the input tensor. The TBNN is composed of two parts. The first one is an usual DNN. The number of the nodes of the final hidden layer of this part is equal to the number of the tensor input layers. The second part of the TBNN is the sum of the tensor input layers multiplied by the outputs of the first part. Then, the symbolic expression of the TBNN is $k^{ij} = \sum_{t} T_t^{ij} \sigma(\sum_a W^L_{ta} \sigma(\sum_b W^{L-1}_{ab} \sigma(\cdots \sum_m W^2\sigma(W^1_{nm}x_m+b^1_n) + \cdots) + b^{L-1}_a) + b^L_k)$, where $k^{ij}$ and $T_t^{ij}$ are the output Eddington tensor and the output of the $t$-th tensor input layer, respectively. Again, a batch normalization layer is inserted just before the final hidden layer of the first part. The TBNN is expected to produce more accurate results than the CWNN because the former respects invariance under the coordinate transformation, which is the most important property of the tensor. The choice of the input tensors depends on the problem and will be explained in the following sections.

The hyperparameters are the model parameters which are not updated by learning but are adjusted to achieve the best performance. In our networks, we have two hyperparameters: the numbers of hidden layers and nodes per a layer. Although the learning rate $\eta$ is also a hyperparameter normally, we fix it to $\eta=10^{-2}$ in this work. We trained the DNNs with different sizes and determined the best values as the one that gives the least validation loss as explained in the next section.

\section{Results} \label{sec:results}
\subsection{Hyperparameter Survey} \label{sec:survey}
As a test problem to determine the hyperparameters, we try to reproduce the M1-Eddington tensor. The M1-Eddington tensor is first evaluated in the fluid-rest frame according to equations (\ref{eq:PM1}--\ref{eq:PM1FRthin}), and then this is converted to the laboratory frame by the Lorentz transformation with the local fluid velocity. The resultant tensor $P_{\rm M1,Lor}^{ij}$ is as follows:
\begin{equation}
P_{\rm M1,Lor}^{ij} = A \gamma^{ij} + B \frac{F_{\rm FR}^i F_{\rm FR}^j}{|F_{\rm FR}|^2} + C \frac{V^iV^j}{|V|^2} + D \frac{F_{\rm FR}^i V^j + F_{\rm FR}^j V^i}{|F_{\rm FR}||V|},
\end{equation}
where the coefficients are given as
\begin{eqnarray}
A &=& \frac{\alpha}{3} E_{\rm FR},\\
B &=& \beta E_{\rm FR},\\
C &=& \left\{ \Gamma^2 E + 2\frac{\Gamma^3}{1+\Gamma} (F_{\rm FR}\cdot V) + \frac{2}{3} \alpha \frac{\Gamma^2}{1+\Gamma}E_{\rm FR} \right. \nonumber \\ &&\left. + \left(\frac{\Gamma^2}{1+\Gamma}\right)^2 E_{\rm FR} \left( \frac{\alpha}{3} |V|^2 + \beta\frac{(F_{\rm FR}\cdot V)^2}{|F_{\rm FR}|^2} \right)\right\} |V|^2, \nonumber \\ && \\
D &=& \left\{ \Gamma + \beta\frac{\Gamma^2}{1+\Gamma}\frac{E_{\rm FR}}{|F_{\rm FR}|^2}(F_{\rm FR}\cdot V) \right\}|F_{\rm FR}||V|,
\end{eqnarray}
and $\Gamma=(1-V^2)^{-1/2}$ is the Lorentz factor.
This is an analytic function of ($E_{\rm FR}$,$F_{\rm FR}^i$,$V^i$). Hence the input data to the networks are ($E_{\rm FR}$,$F_{\rm FR}^i$,$V^i$). The tensor basis that show up in this equation exhausts possible combinations of the unit tensor, energy flux, and velocity for the symmetric tensor. The tensor input layers for the TBNN are hence $F_{\rm FR}^i F_{\rm FR}^j/|F_{\rm FR}|^2$, $\gamma^{ij}$, $V^i V^j/|V|^2$, and $(F_{\rm FR}^i V^j+F_{\rm FR}^j V^i)/|F_{\rm FR}||V|$. The output is the $k_{\rm M1,Lor}^{ij}:=P_{\rm M1,Lor}^{ij}/E_{\rm M1,Lor}$, with $E_{\rm M1,Lor} = \Gamma^2 E_{\rm FR} + 2\Gamma^2 (F_{\rm FR}\cdot V)+(\alpha \Gamma^2/3 +\beta\Gamma (F_{\rm FR}\cdot V)/|F_{\rm FR}|^2) E_{\rm FR}|V|^2$ being the Lorentz-transformed energy density.

\begin{figure}[t]
\plotone{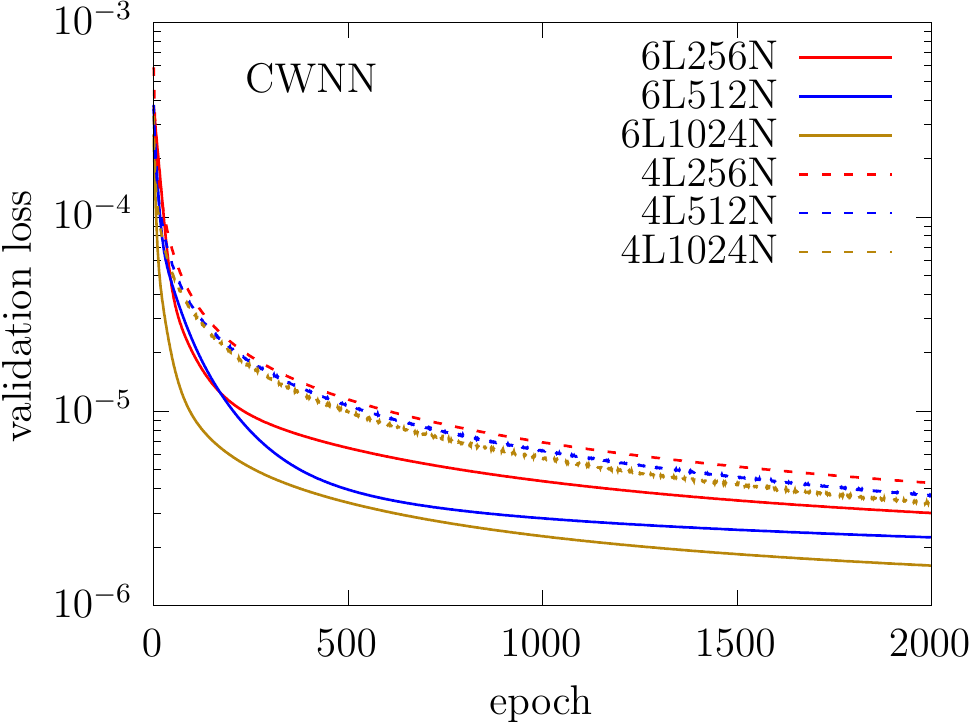} \\
\plotone{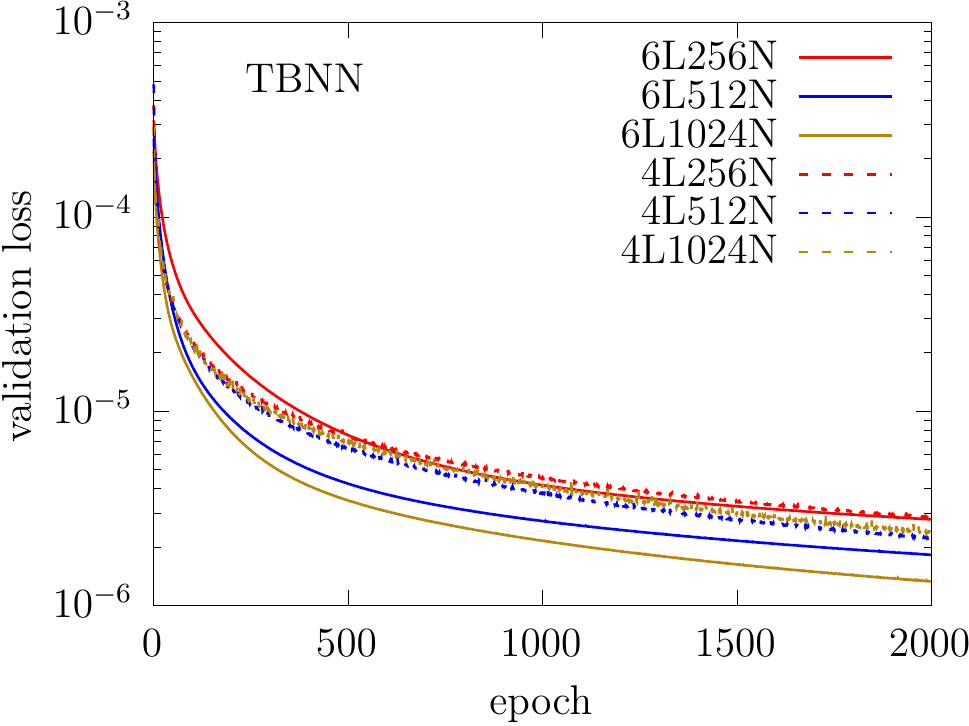}
\caption{The learning curves of the validation loss for different networks with different hyperparameters. The upper panel shows the validation losses for the CWNN, while the lower exhibits those for the TBNN. The numbers of hidden layers and nodes are indicated in the legend by digits before `L' and `N', respectively: the red, blue, and dark-yellow curves represent the results for the $256$-, $512$-, and $1024$-node networks; the solid and dashed lines correspond to the $6$- and $4$-hidden-layer networks, respectively. \label{fig:validationloss}}
\end{figure}

Figure \ref{fig:validationloss} shows the learning curves of the validation loss for different hyperparameters. The learning curve is the loss as a function of epoch. Here, the hyperparameters are chosen as follows: the number of hidden layers is $4$ or $6$; the number of nodes is $256$, $512$, or $1024$. Note that these parameters correspond to the hidden layers before the batch normalization layer shown in figures \ref{fig:CWNN} and \ref{fig:TBNN}, i.e., the final hidden layer, which outputs the Eddington tensor or the coefficients of the tensor input, is not counted. For this test, we continue learning for $2000$ epochs. As shown in the figure, the network with $6$ hidden layers and $1024$ nodes performed best, i.e., the validation loss is the smallest at the final epoch. Although it is not shown here, the training loss is also decreasing. We hence decided to use the networks with these sizes in the remainder of this paper.

\begin{figure}[t]
\plotone{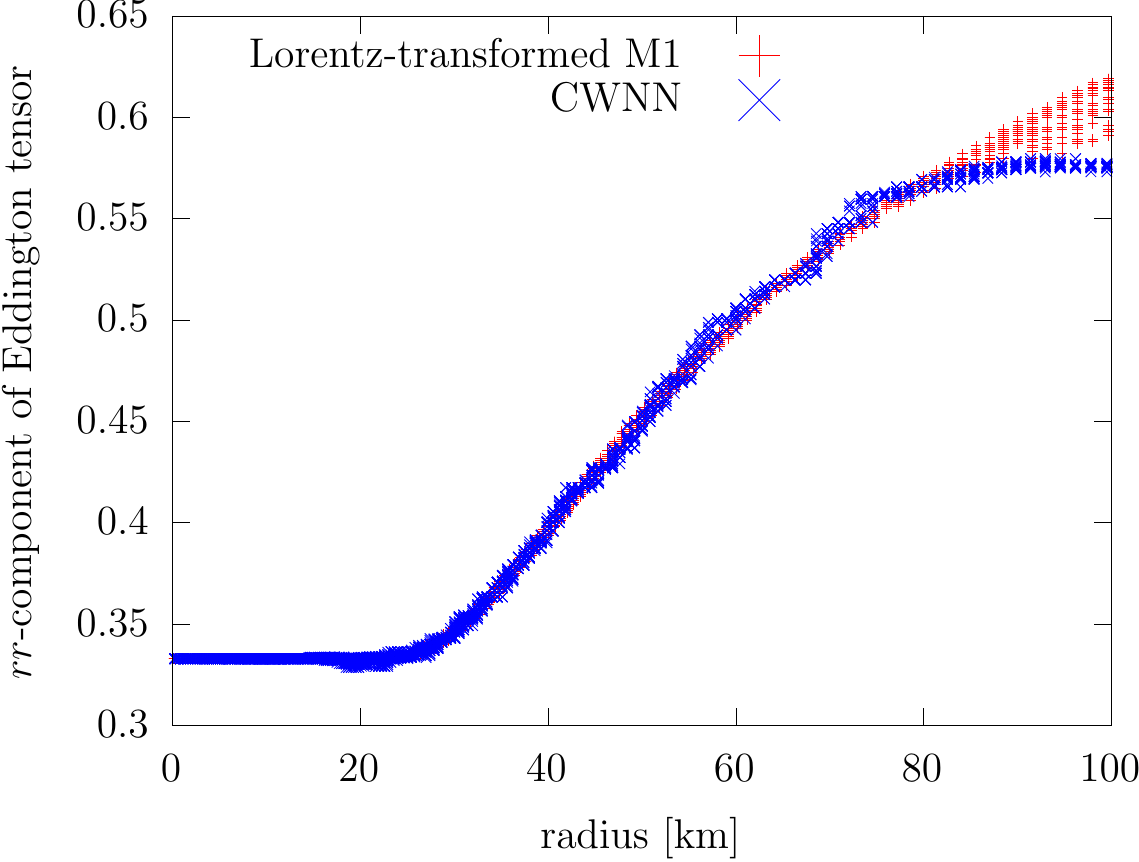}
\caption{The radial profiles of $k_{\rm M1,Lor}^{rr}$ (red) and the output counterpart of the CWNN (blue) at $\epsilon = 1\,{\rm MeV}$. Although the data are distributed in the $r$--$\theta$ plain originally, they are $\theta$-marginalized here. \label{fig:LorCWNN}}
\end{figure}

Figure \ref{fig:LorCWNN} shows the output of the CWNN and its supervisor data. We show only the $rr$-component of the Eddington tensor at the neutrino energy of $\epsilon=1\,{\rm MeV}$ here. The output of the CWNN reproduces $k_{\rm M1,Lor}^{rr}$ at $r\la 80\,{\rm km}$ quite well. The reason for the deviation seen at $r\ga80\,{\rm km}$ is as follows: the number of samples with large values of the $rr$-component is relatively small because only low-energy neutrinos have strongly forward-peaked distributions owing to lower matter opacities to them at large radii in the domain where we sampled the data. If we enlarge the domain to obtain more samples, the discrepancy should get smaller.

\subsection{Deep Learning of the Boltzmann-Eddington tensor}
Using the data derived from the Boltzmann-radiation-hydrodynamics simulation and the hyperparameters determined in section \ref{sec:survey}, we trained the CWNN and TBNN for $10000$ epochs. In the following, we show the accuracy of the closure relations so obtained from the CWNN and TBNN in sections \ref{sec:cwnn} and \ref{sec:tbnn}, respectively. We also compare them with the M1-closure relation.

\subsubsection{Component-wise Neural Network} \label{sec:cwnn}
\begin{figure}[t]
\plotone{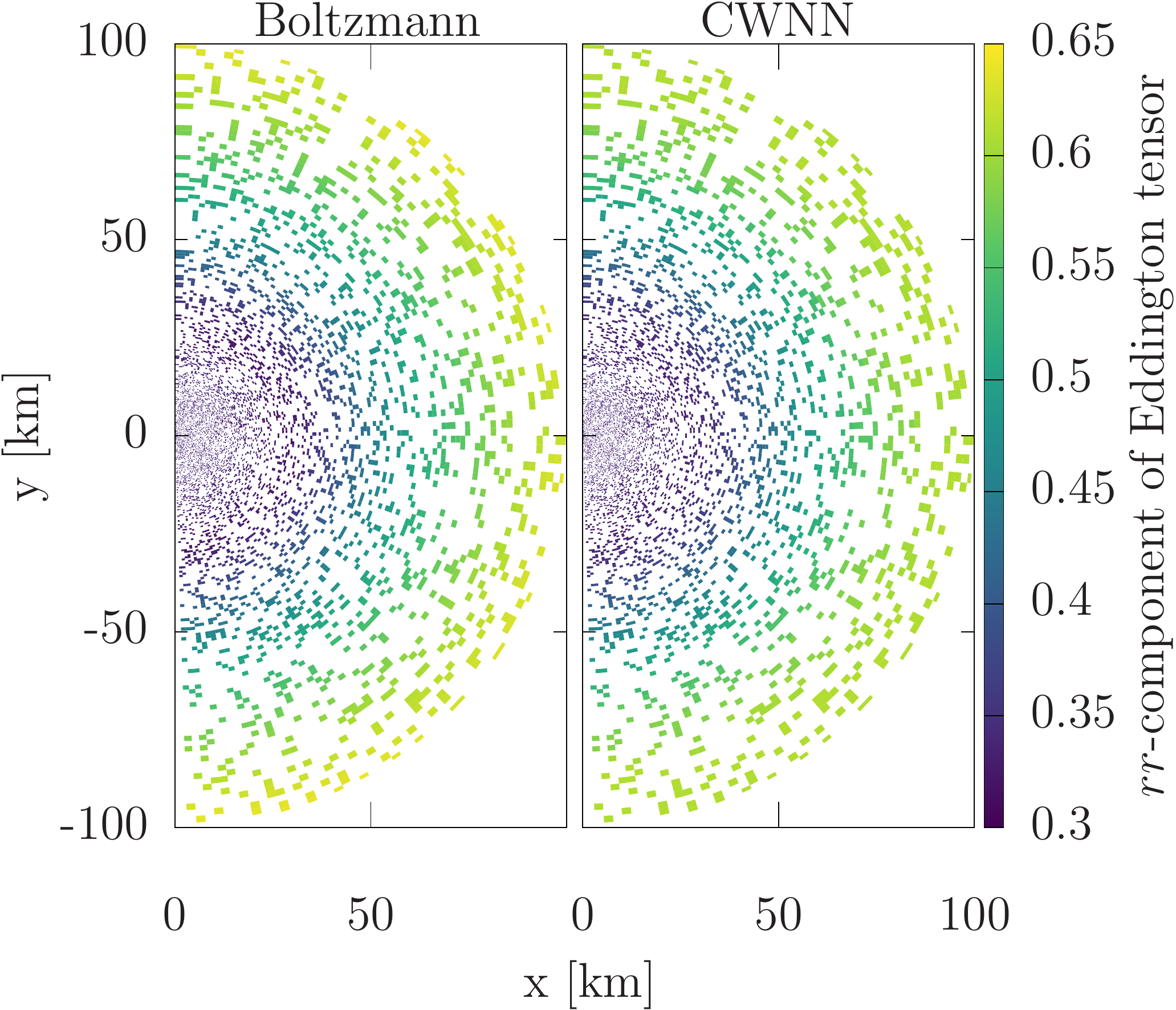}
\caption{The spatial distribution of values of the $rr$-components of the Boltzmann- ($k_{\rm Boltz,LB}^{rr}$, left) and CWNN- ($k_{\rm CWNN}^{rr}$, right) Eddington tensors for $\epsilon = 1\,{\rm MeV}$. \label{fig:kmapCWNN}}
\end{figure}

Figure \ref{fig:kmapCWNN} shows again the spatial distribution of the $rr$-component for the CWNN-Eddington tensor $k_{\rm CWNN}^{rr}$. Here, $k_{\rm CWNN}^{ij}$ is the Eddington tensor obtained with the CWNN. In the training, the input data are $E_{\rm LB}$, $F_{\rm LB}^i$, and $V^i$, and the output supervisor is the Boltzmann-Eddington tensor $k_{\rm Boltz,LB}^{ij}$. This figure also presents the supervisor for comparison. After completing the learning procedure for the training set, we apply the network to the validation set to obtain the CWNN-Eddington tensor $k_{\rm CWNN}^{ij}$ as the output of the network for the input data $(E_{\rm LB}, F_{\rm LB}^i, V^i)$. Figure \ref{fig:kmapCWNN} presents $k_{\rm CWNN}^{rr}$ thus obtained; the plot is a bit sporadical, since the validation set consists of just $20\%$ of the entire data; note also that only the data for the lowest neutrino-energy bin at $\epsilon=1\,{\rm MeV}$ is shown here.

\begin{figure}[t]
\plotone{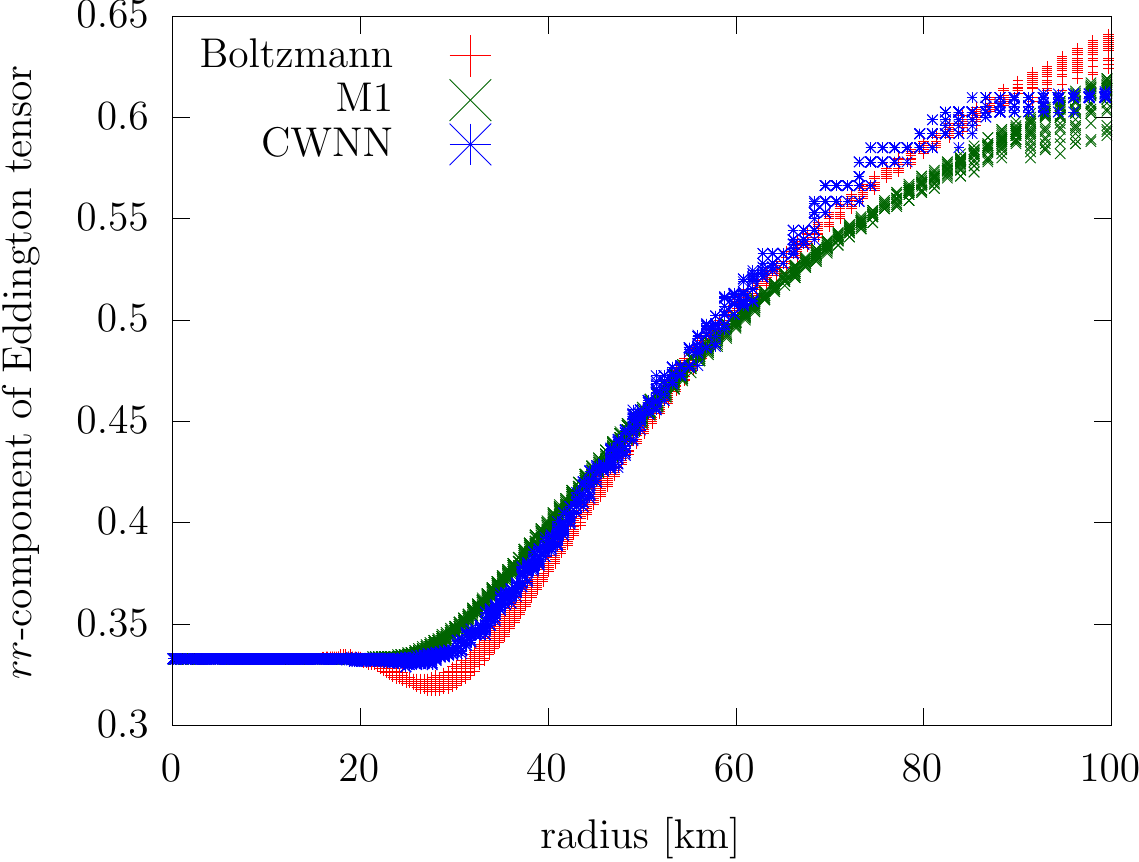}
\caption{The radial distributions of values of the $rr$-components of the Eddington tensors at $\epsilon = 1\,{\rm MeV}$ for the validation set. The data are $\theta$-marginalized. The red, green, and blue dots represent the Boltzmann-, M1-, and CWNN-Eddington tensors, respectively. \label{fig:CWNN_dot_low}}
\end{figure}

We collect $k_{\rm CWNN}^{rr}$'s in the same radial bins and plot the radial distribution in figure \ref{fig:CWNN_dot_low}. The figure also presents the radial distributions of $k_{\rm Boltz,LB}^{rr}$ and $k_{\rm M1,LB}^{rr}$, where $k_{\rm Boltz,LB}^{ij}$ is the actual Eddington tensor while $k_{\rm M1,LB}^{ij}$ is calculated from the same input data ($E_{\rm LB}$, $F_{\rm LB}^i$, $V^i$) through equations (\ref{eq:PM1}--\ref{eq:beta}, \ref{eq:PM1LBthick}, \ref{eq:PM1LBthin}). The data points have some scatter in each radial bin, since the Eddington tensors are not uniform with respect to the angle $\theta$. Some trend is apparent in the figure, however: the $rr$-component of the Eddington tensor is approximately equal to the Eddington factor and takes the value of $1/3$ in the inner optically thick region and increases to unity in the outer optically thin region.

\begin{figure*}[t]
\centering
\begin{tabular}{ccc}
\includegraphics[width=0.3\hsize]{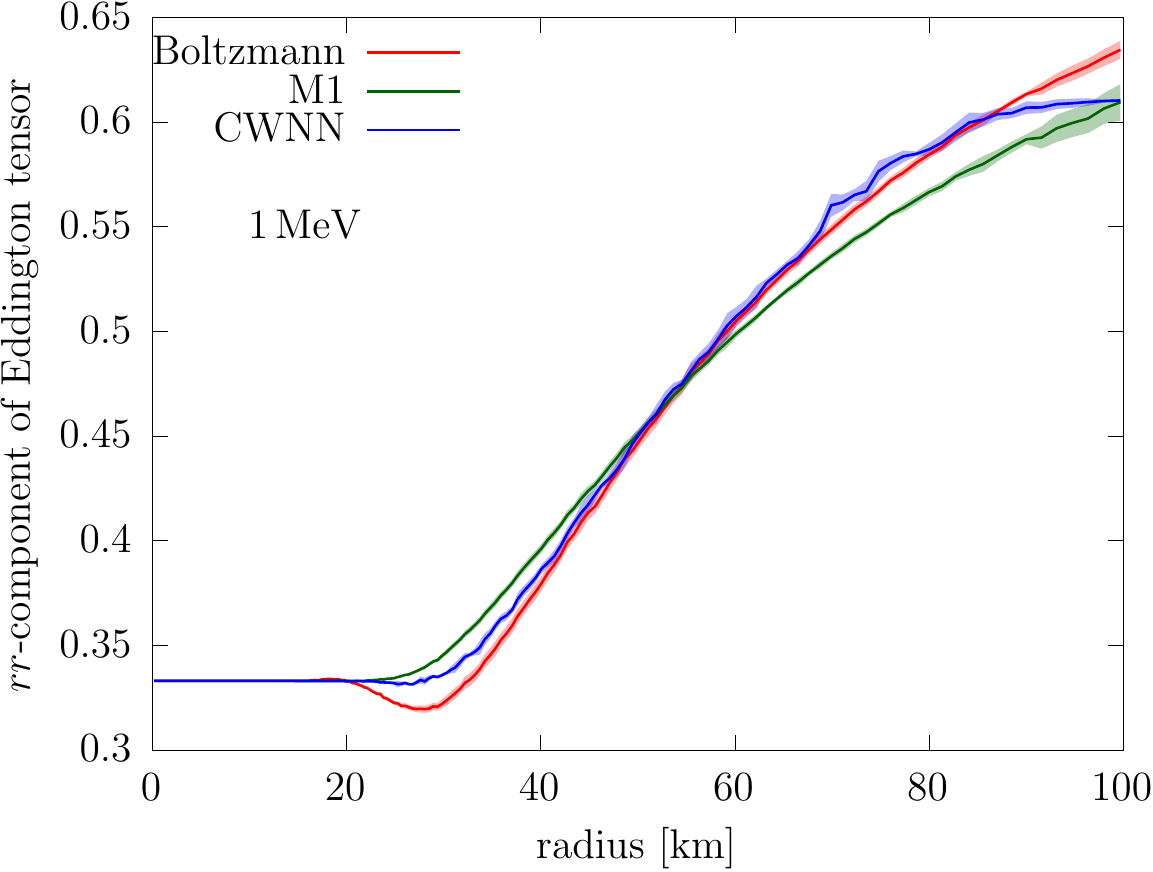} & \includegraphics[width=0.3\hsize]{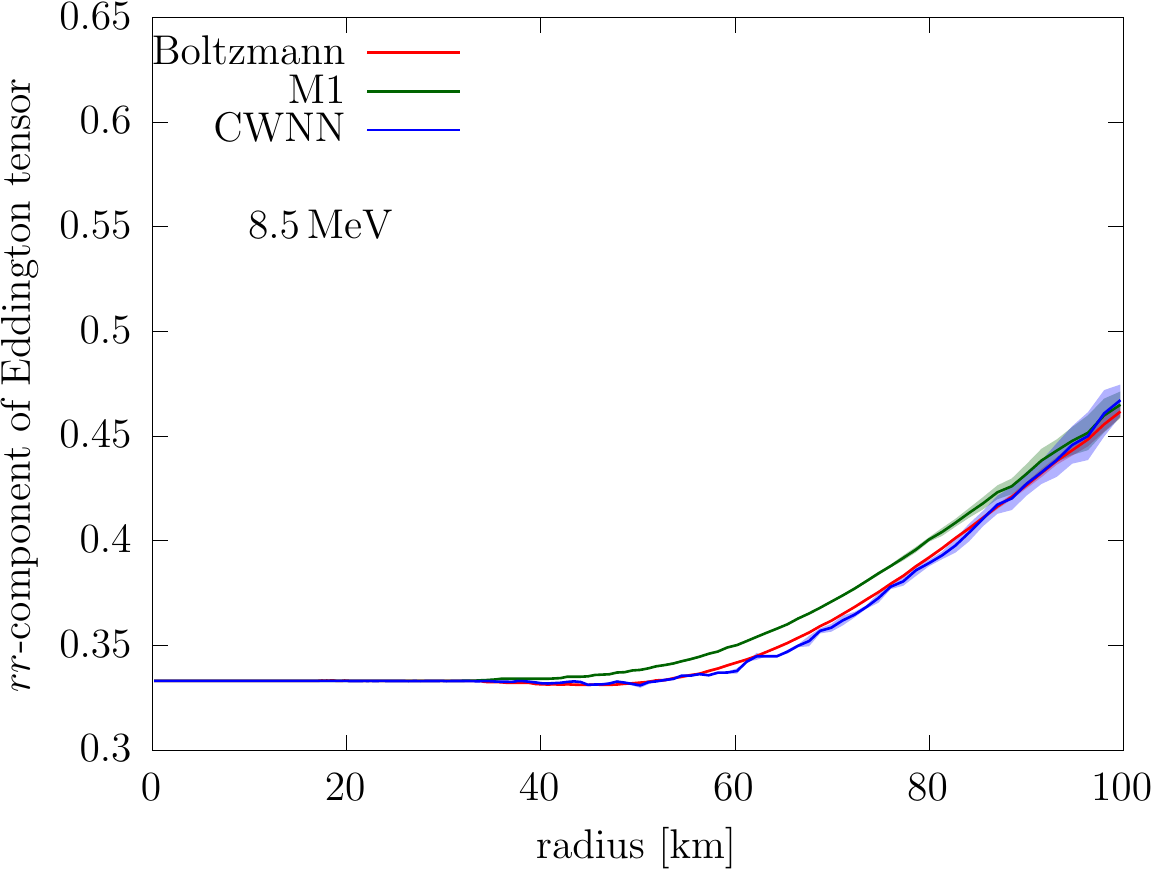} & \includegraphics[width=0.3\hsize]{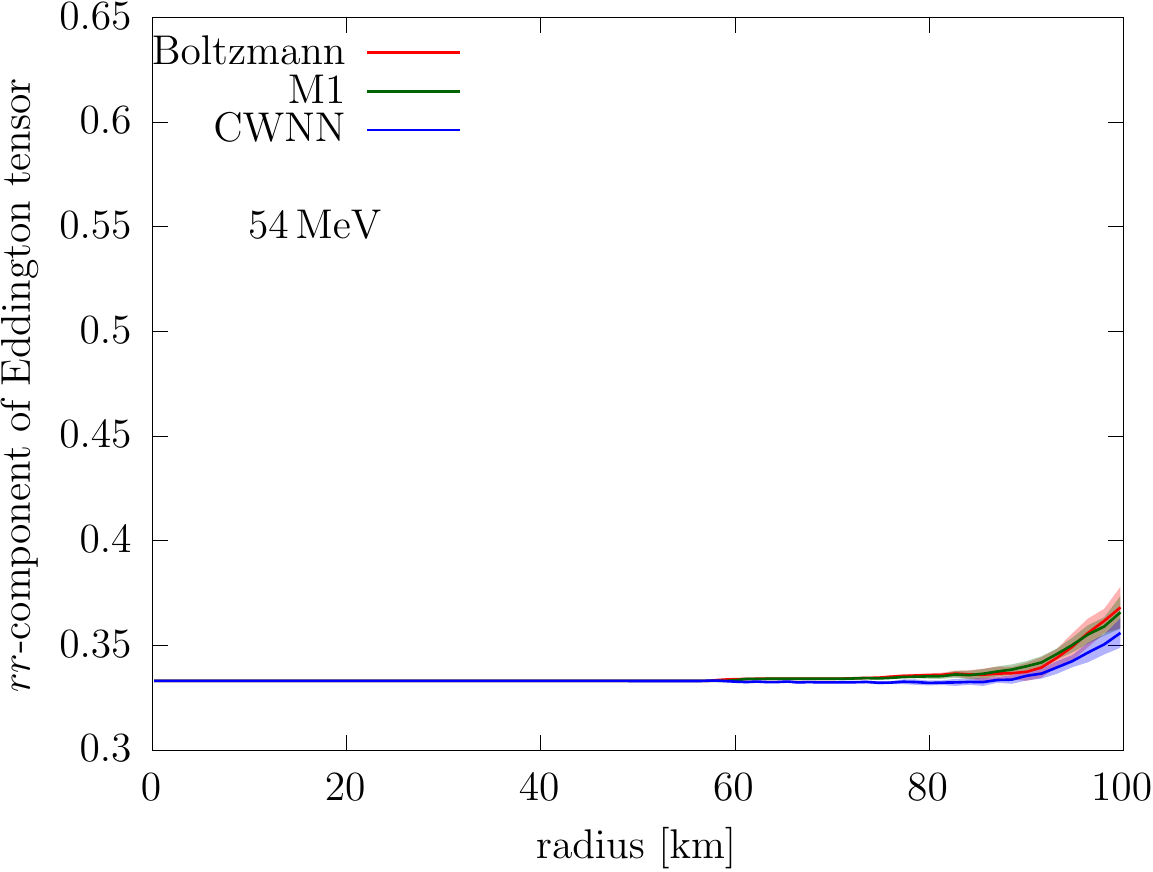} 
\end{tabular}
\caption{The mean and standard deviation of the angular distribution of the $rr$-components of the Eddington tensors as a function of the radius. The solid lines indicate the mean values, and the standard deviations from these mean values are displayed by shaded areas. The red, green, and blue colors correspond to the Boltzmann-, M1-, and CWNN-Eddington tensors, respectively. The left, middle, and right panels show the results for neutrinos with $\epsilon=1\,{\rm MeV}$, $8.5\,{\rm MeV}$, and $54\,{\rm MeV}$, respectively. \label{fig:CWNN_fill}}
\end{figure*}

In order to clarify the trend and the scatter from it quantitatively, figure \ref{fig:CWNN_fill} shows the radial distributions of the mean values as well as the standard deviations. The mean is actually the average over the angle $\theta$ with the weight equal to the $\theta$-width of the bin; the bins which do not have the data such as the blank cells in figure \ref{fig:kmapCWNN} are excluded from the summation for the mean and weight normalization. The standard deviation is defined in the same way.

The left panel of figure \ref{fig:CWNN_fill} displays again the results for the neutrinos with $\epsilon=1\,{\rm MeV}$. The Boltzmann-Eddington tensor takes $1/3$ in the innermost regions and increases in the outer regions. In the regions with $20\,{\rm km}\la r \la 40\,{\rm km}$, however, it is smaller than $1/3$ as seen in, e.g., \citet{1992A&A...256..452J, 2020ApJ...903...82I}. This is because the opacity of neutrinos rapidly decreases with radius there. It is like a radiation from a sphere with a constant brightness: outgoing neutrinos are almost isotropic, while ingoing neutrinos are essentially lacked; then the netrino distribution is approximately hemispheric in fact with the Eddington factor less than $1/3$. Neither of the M1- nor CWNN-Eddington tensors reproduces this trend well. This happens for the M1-Eddington tensor because it is designed from the beginning to take a value in between $1/3$ and $1$. The reason for the CWNN-Eddington tensor is that it is influenced by the behavior in other energy bins; this point will be discussed below. In the region $40\,{\rm km}\la r \la 80\,{\rm km}$, the CWNN-Eddington tensor tracks the Boltzmann-Eddington tensor more closely than the M1-Eddington tensor. This is one of the achievements of our machine learning strategy.

The middle panel of figure \ref{fig:CWNN_fill} shows the Eddington tensors for the neutrinos with $\epsilon=8.5\,{\rm MeV}$. This energy is close to the mean energy at the shock radius. The CWNN-Eddington tensor follows the Boltzmann-Eddington tensor very closely. The M1-Eddington tensor, on the other hand, has some deviations in the region $40\,{\rm km} \la r \la 90\,{\rm km}$. It is worth noting that the trough structure, i.e., the Eddington tensor less than $1/3$ as seen in the left panel, is far less significant in this case. This is because the opacity for this energy bin crosses $2/3$ more gradually and, as a result, the angular distribution varies from the isotropic one to the forwardly peaked one without becoming hemispheric in between.

Although the appearance of the trough structure depends on the neutrino energy, our machine learning architecture does not learn it well and choose to produce a small dip at all energies to reduce the loss function. This is the reason why the CWNN-Eddington tensor in the left panel fails to reproduce the deep trough structure found in the Boltzmann-Eddington tensor. On the other hand, the shallow trough in the middle panel is successfully reproduced by our CWNN.

The right panel of figure \ref{fig:CWNN_fill} represents the Eddington tensors for the neutrinos with $\epsilon=54\,{\rm MeV}$. Owing to the high energy, the opacity is large and the $rr$-component of the Eddington tensor is close to $1/3$, the value for the isotropic case, up to large radii. In the region $80\,{\rm km}\la r \la 100\,{\rm km}$, the M1-Eddington tensor rather than the CWNN-Eddington tensor is close to the Boltzmann-Eddington tensor. This is again because the CWNN-Eddington tensor produces a shallow trough structure, which should be absent in this case. This also leads to a delay in the rise of $k_{\rm CWNN}^{rr}$. The M1-Eddington tensor, on the other hand, does not generate a trough and, as a result, well traces the Boltzmann-Eddington tensor in this particular case.

Next, we shift our attention to the off-diagonal $r\theta$-component of the Eddington tensors. The main concern about the $rr$-component was whether it follows the well-known behavior of the Eddington factor or not. This is the reason why we considered the mean value of the $rr$-component in figure \ref{fig:CWNN_fill}. On the other hand, the values of the $r\theta$-component are distributed around zero, and hence the main focus is whether the $r\theta$-component of the CWNN-Eddington tensor traces correctly the deviation of the Boltzmann-Eddington tensor from zero. In the following, we consider the differences of the CWNN- and M1-Eddington tensors from the Boltzmann-Eddington tensor.

\begin{figure}[t]
\plotone{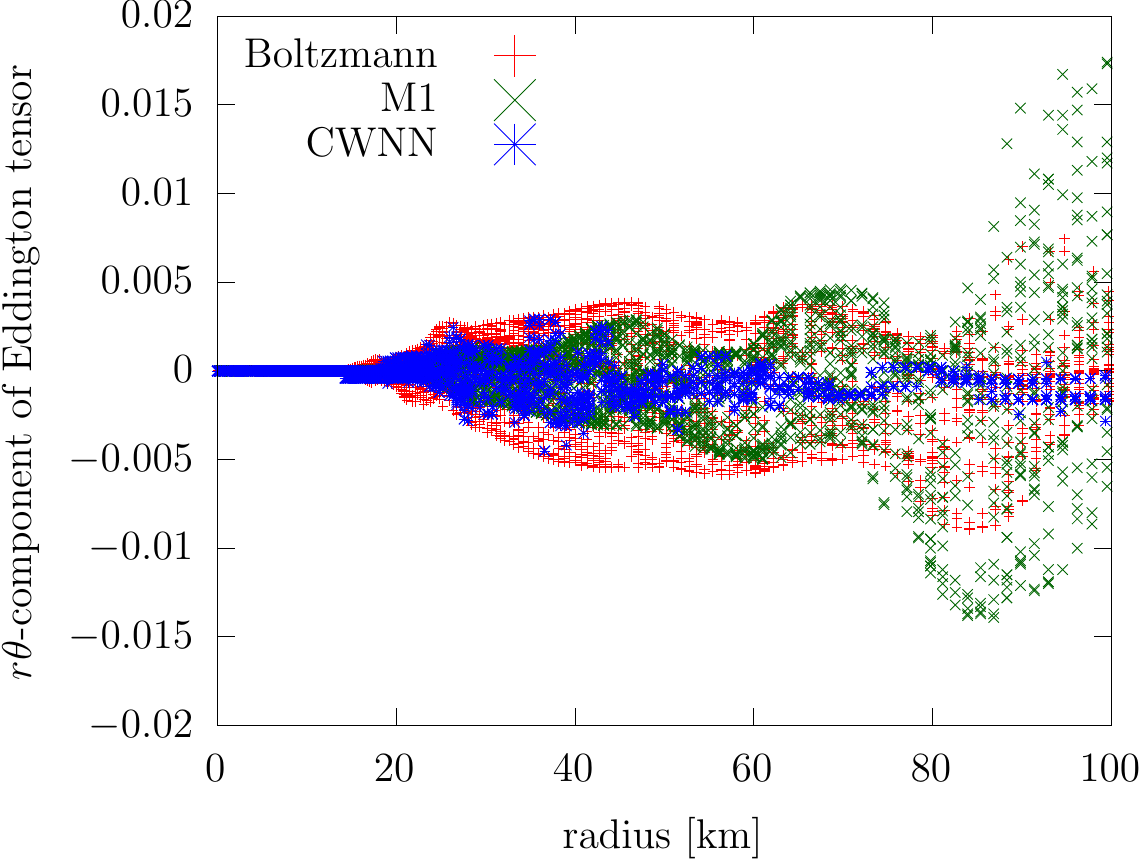}\\
\plotone{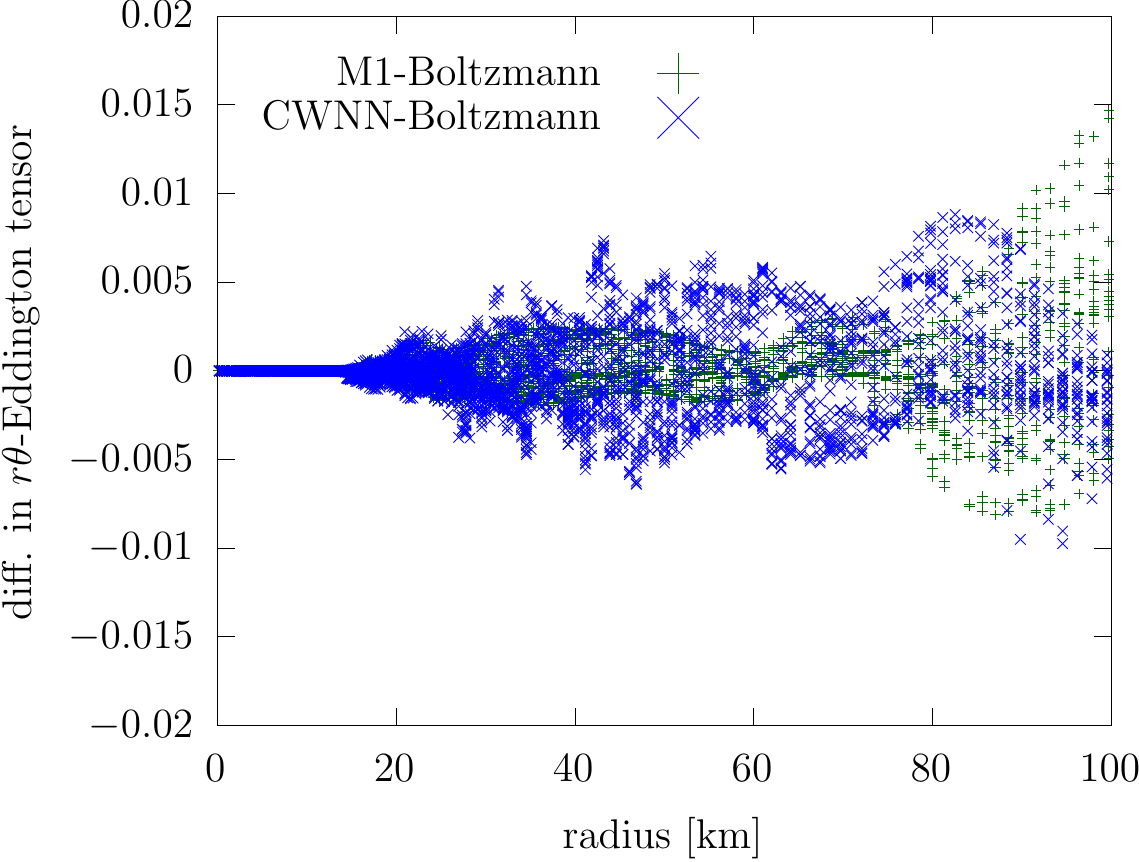}
\caption{The radial distributions of the $r\theta$-components of the Eddington tensors with $\epsilon=1\,{\rm MeV}$. The upper panel shows the raw data; the red, green, and blue dots represent the Boltzmann-, M1-, and CWNN-Eddington tensors, respectively. The lower panel shows the difference of the M1- (green) and CWNN- (blue) Eddington tensors from the Boltzmann-Eddington tensor. \label{fig:CWNNrth_dot_low}}
\end{figure}

\begin{figure*}[t]
\centering
\begin{tabular}{ccc}
\includegraphics[width=0.3\hsize]{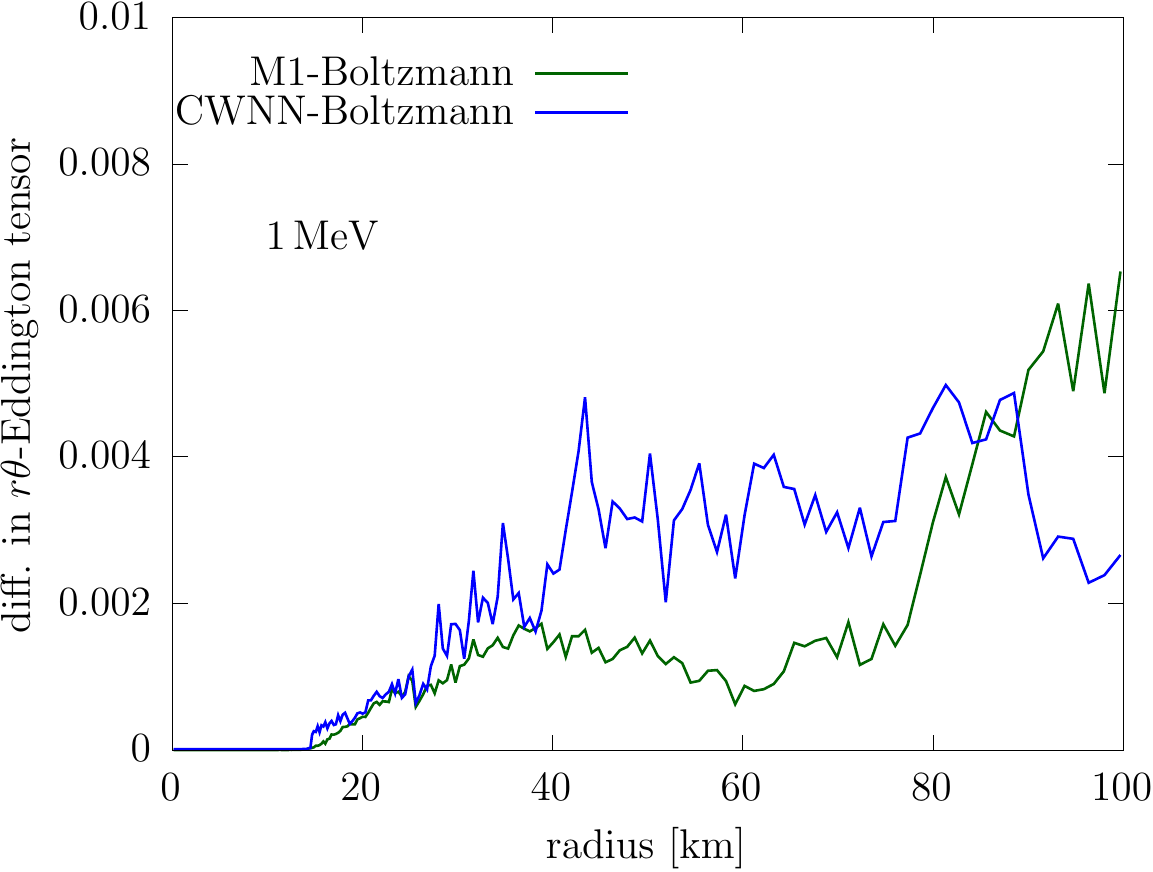} & \includegraphics[width=0.3\hsize]{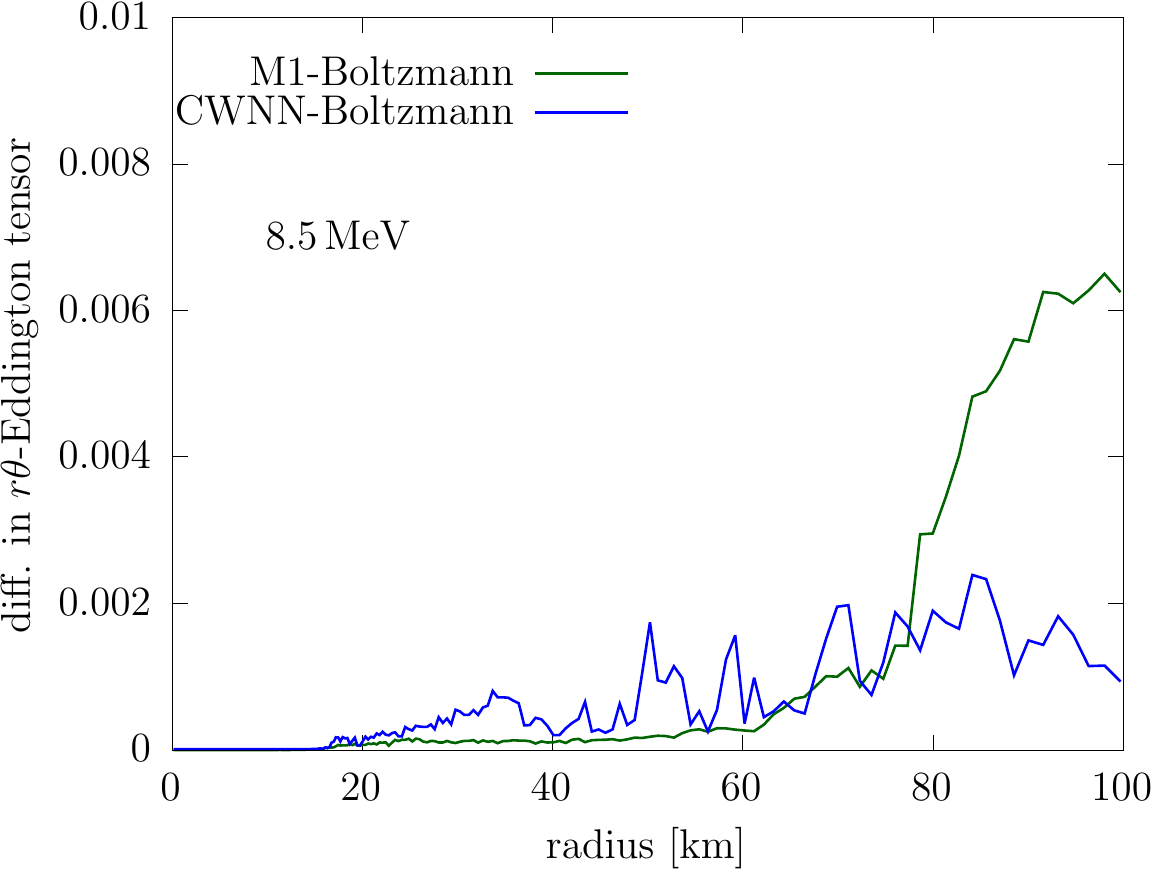} & \includegraphics[width=0.3\hsize]{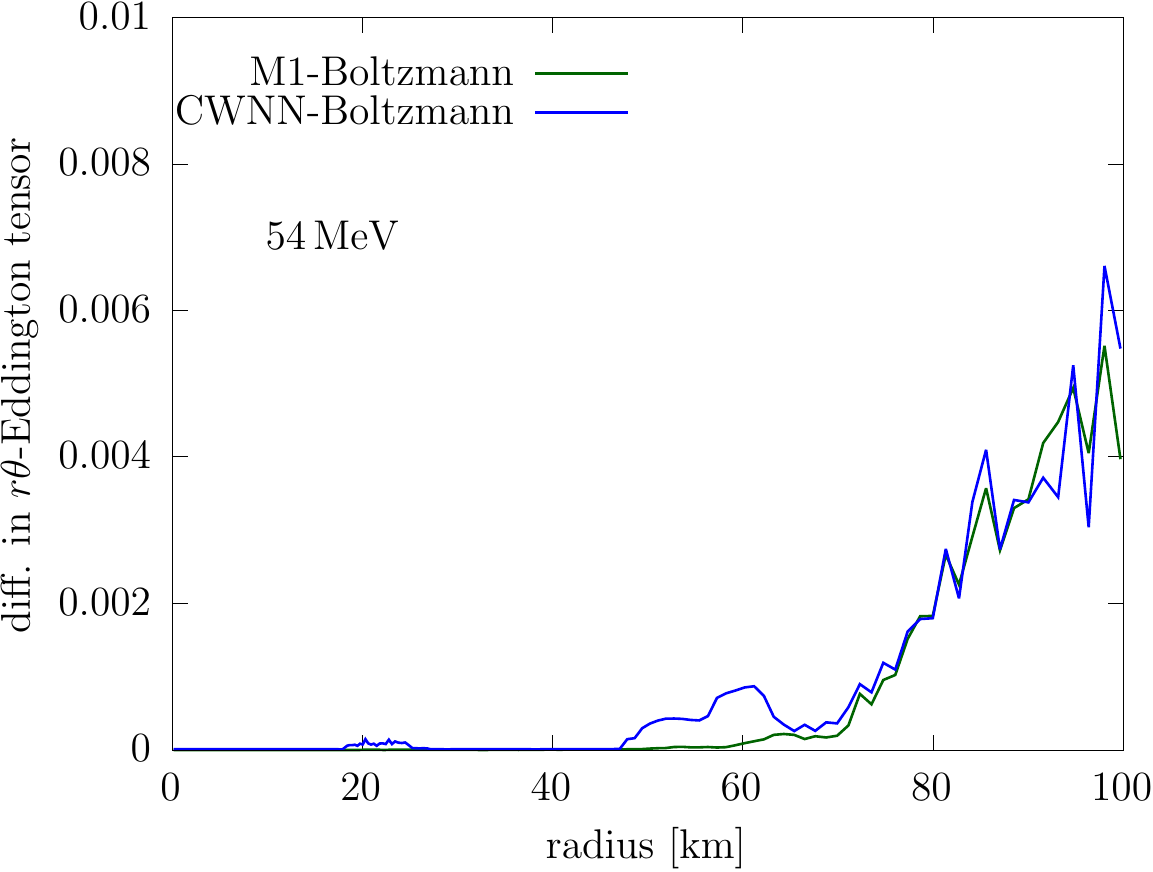} 
\end{tabular}
\caption{The radial profiles of the rms differences between the $r\theta$-components of the Eddington tensors. The green lines show the difference between the M1- and Boltzmann-Eddington tensors, whereas the blue lines present the differences between the CWNN- and Boltzmann-Eddington tensors. The left, middle, and right panels correspond to the neutrino energies of $\epsilon=1\,{\rm MeV}$, $8.5\,{\rm MeV}$, and $54\,{\rm MeV}$, respectively. \label{fig:CWNNrth_diffline}}
\end{figure*}

Figure \ref{fig:CWNNrth_dot_low} shows the comparison for the raw data. The upper panel exhibits the component themselves, while the lower panel provides the deviations from the Boltzmann-Eddington tensor. It is observed that the scatter is larger for the CWNN-Eddington tensor than for the M1-Eddington tensor. For more quantitative discussions, we consider the root-mean squares (rms) of these data. They are shown in figure \ref{fig:CWNNrth_diffline} for some energy bins.

The left panel of figure \ref{fig:CWNNrth_diffline} is for $\epsilon=1\,{\rm MeV}$, corresponding to the lower panel of figure \ref{fig:CWNNrth_dot_low}. The rms difference between the M1- and Boltzmann-Eddington tensors is similar to that between the CWNN- and Boltzmann-Eddington tensors at $r\la 40\,{\rm km}$. In the region $40\,{\rm km} \la r \la 80\,{\rm km}$, the former is about half of the latter. At $r\ga 80\,{\rm km}$, the difference between the M1- and Boltzmann-Eddington tensors increases suddenly to $0.006$, while the difference between the CWNN- and Boltzmann-Eddington tensors remains $\sim 0.004$ or even decreases. Since the velocity is large in this region, it seems that the influence of the velocity on the Eddington tensor in the laboratory frame is treated better by the CWNN than by the M1-closure scheme.

The middle and right panels of figure \ref{fig:CWNNrth_diffline} represent the results for $\epsilon=8.5\,{\rm MeV}$ and $54\,{\rm MeV}$, respectively. In the middle panel, the M1-Eddington tensor traces the Boltzmann-Eddington tensor well inside $80\,{\rm km}$, and the CWNN-Eddington tensor is as good as or slightly worse than the M1-Eddington tensor. Outside $80\,{\rm km}$, the M1-Eddington tensor again deviates from the Boltzmann-Eddington tensor quickly, while the CWNN-Eddington tensor is still in good agreement with the Boltzmann-Eddington tensor. In the right panel, the rms differences of both tensors are very similar to each other. The CWNN-Eddington tensor does not follow the Boltzmann-Eddington tensor very accurately at this high neutrino energy. This is true in fact for even higher energies: at the highest energy considered in our learning, $\epsilon = 300\,{\rm MeV}$, the agreement of the Eddington tensors is even worse. 

\subsubsection{Tensor Basis Neural Network} \label{sec:tbnn}
We now turn to the TBNN-Eddington tensor $k_{\rm TBNN}^{ij}$ which is the output for the validation set of the trained TBNN. The tensor input layers here are $F_{\rm LB}^i F_{\rm LB}^j/|F_{\rm LB}|^2$, $\gamma^{ij}$, $V^i V^j/|V|^2$, and $(F_{\rm LB}^i V^j+F_{\rm LB}^j V^i)/|F_{\rm LB}||V|$, and hence the number of nodes in the final hidden layer is four. The network structure is the same as that employed in section \ref{sec:survey}, but the input quantities are different. Note that the antisymmetric tensor $(F_{\rm LB}^i V^j - F_{\rm LB}^j V^i)/|F_{\rm LB}||V|$ is not considered because the Eddington tensor is symmetric.

\begin{figure*}[t]
\centering
\begin{tabular}{ccc}
\includegraphics[width=0.3\hsize]{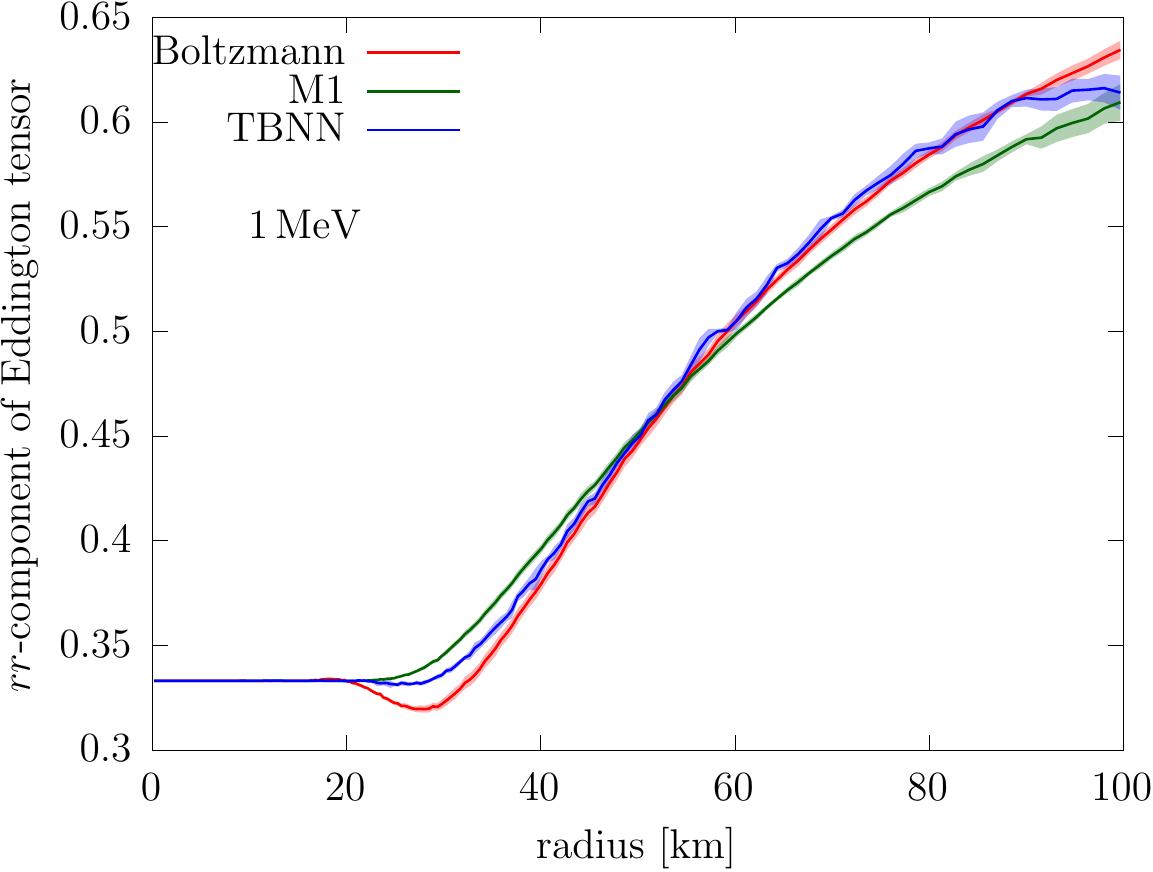} & \includegraphics[width=0.3\hsize]{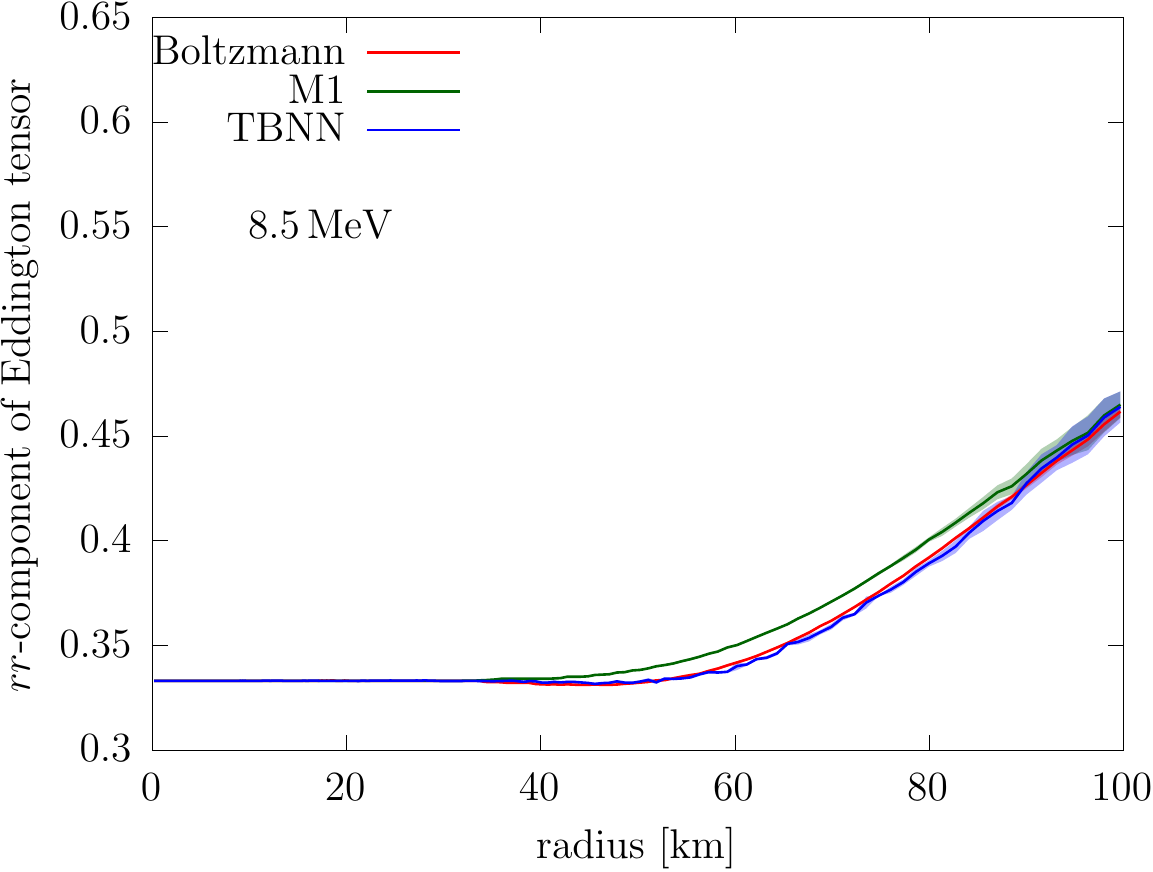} & \includegraphics[width=0.3\hsize]{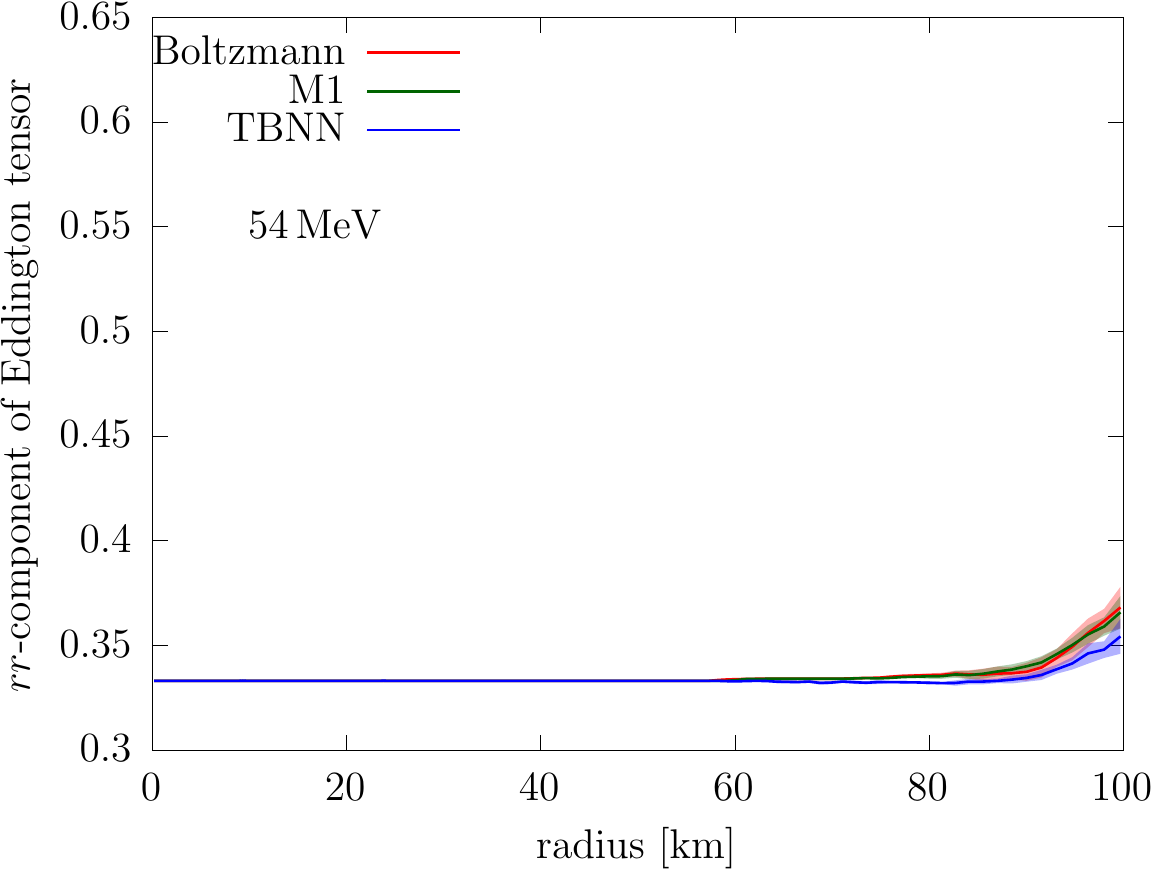} 
\end{tabular}
\caption{The same as figure \ref{fig:CWNN_fill} except that the TBNN-Eddington tensor is shown. \label{fig:TBNN_fill}}
\end{figure*}

Since they are very similar, we skip the plots corresponding to figures \ref{fig:kmapCWNN} and \ref{fig:CWNN_dot_low}. Instead, we show in figure \ref{fig:TBNN_fill} the mean and deviations of the $rr$-components of the TBNN-, M1-, and Boltzmann-Eddington tensors over the $\theta$-coordinates.

The behavior of the TBNN-Eddington tensor is similar to that of the CWNN-Eddington tensor: although the deep trough structure around $r\sim 30\,{\rm km}$ is not reproduced well, it traces the Boltzmann-Eddington tensor better than the M1-Eddington tensor at the low neutrino energy of $1\,{\rm MeV}$; at $8.5\,{\rm MeV}$, roughly the mean energy at the shock radius, the Boltzmann-Eddington tensor is reproduced very well; the agreement gets worse than the M1-Eddington tensor at the high energy of $54\,{\rm MeV}$. Note that the TBNN performs slightly better than the CWNN at $r \ga 80\,{\rm km}$ for the low-energy neutrinos. As discussed in section \ref{sec:survey}, the number of samples there is small, and the learning is difficult. Because the TBNN achieves better results despite such small samples, we judge that it has a slightly better ability to reproduce the Boltzmann-Eddington tensor.

\begin{figure*}[t]
\centering
\begin{tabular}{ccc}
\includegraphics[width=0.3\hsize]{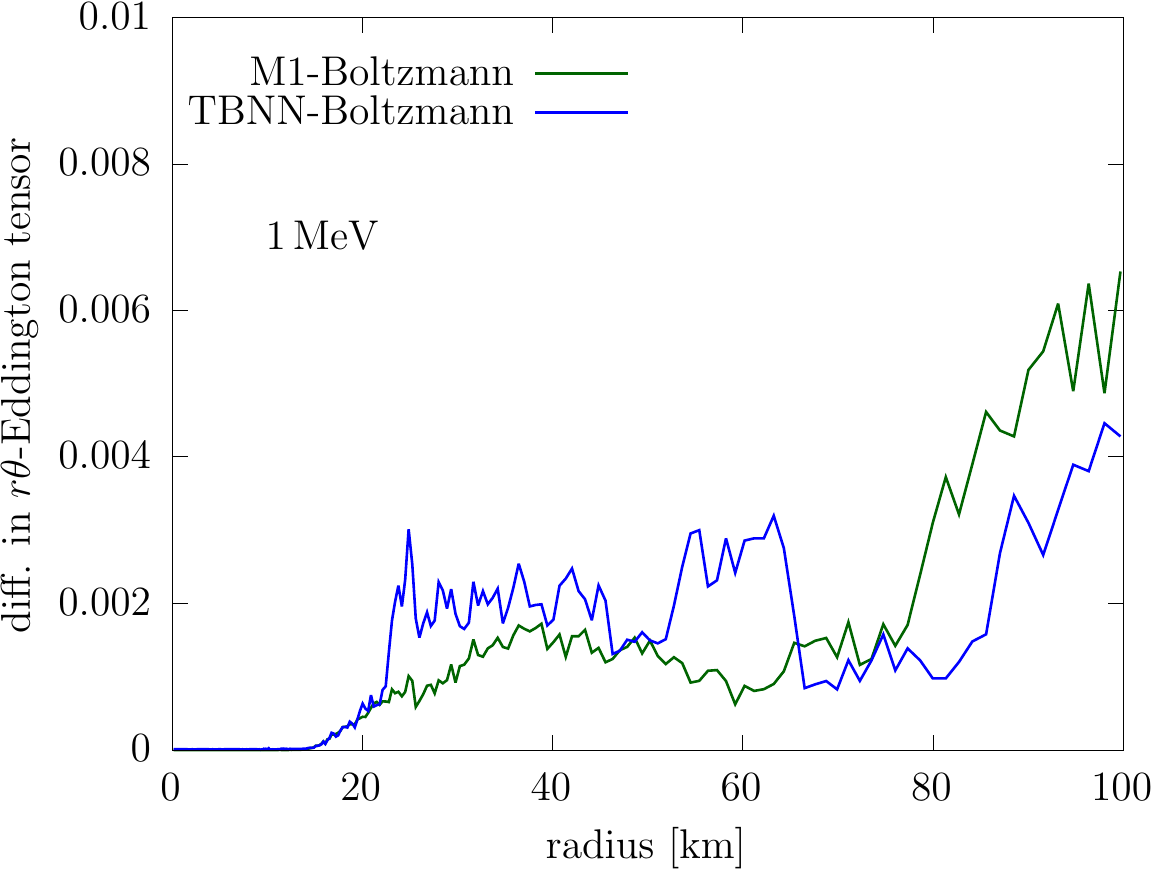} & \includegraphics[width=0.3\hsize]{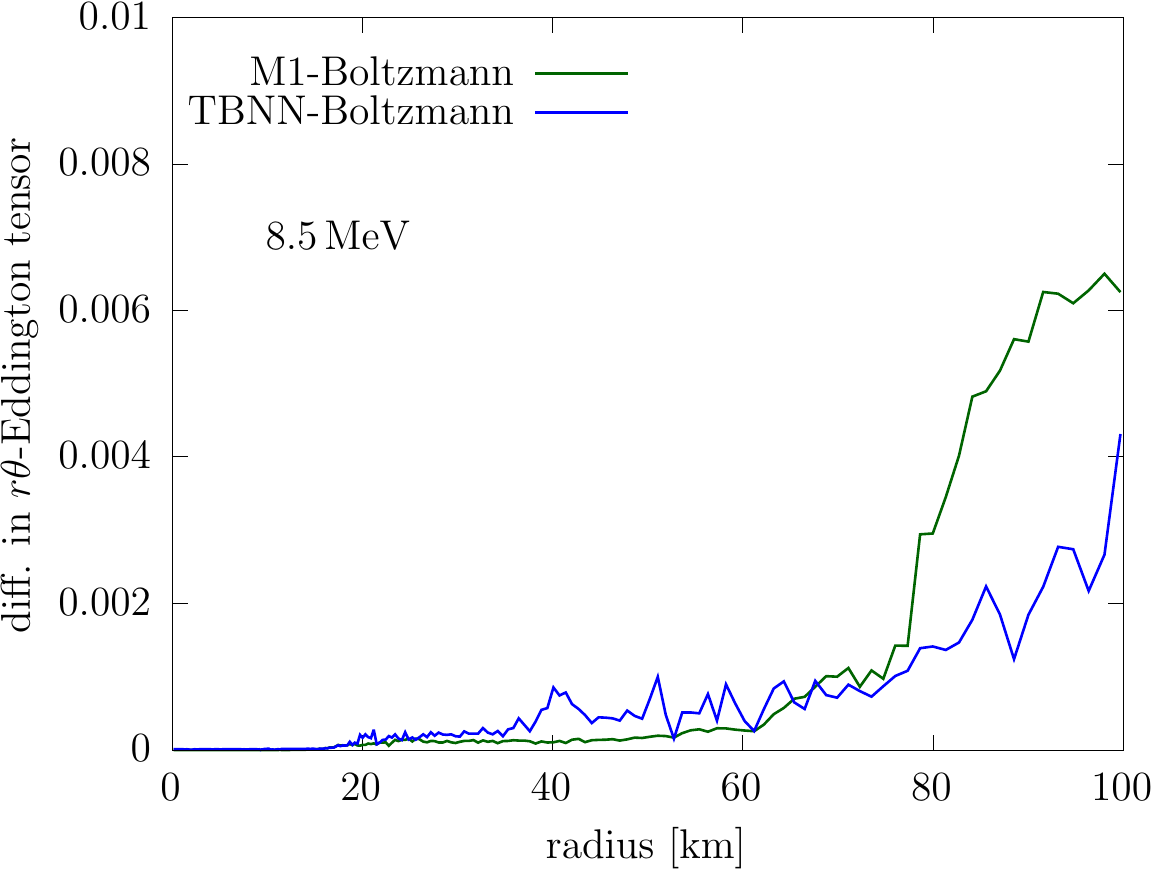} & \includegraphics[width=0.3\hsize]{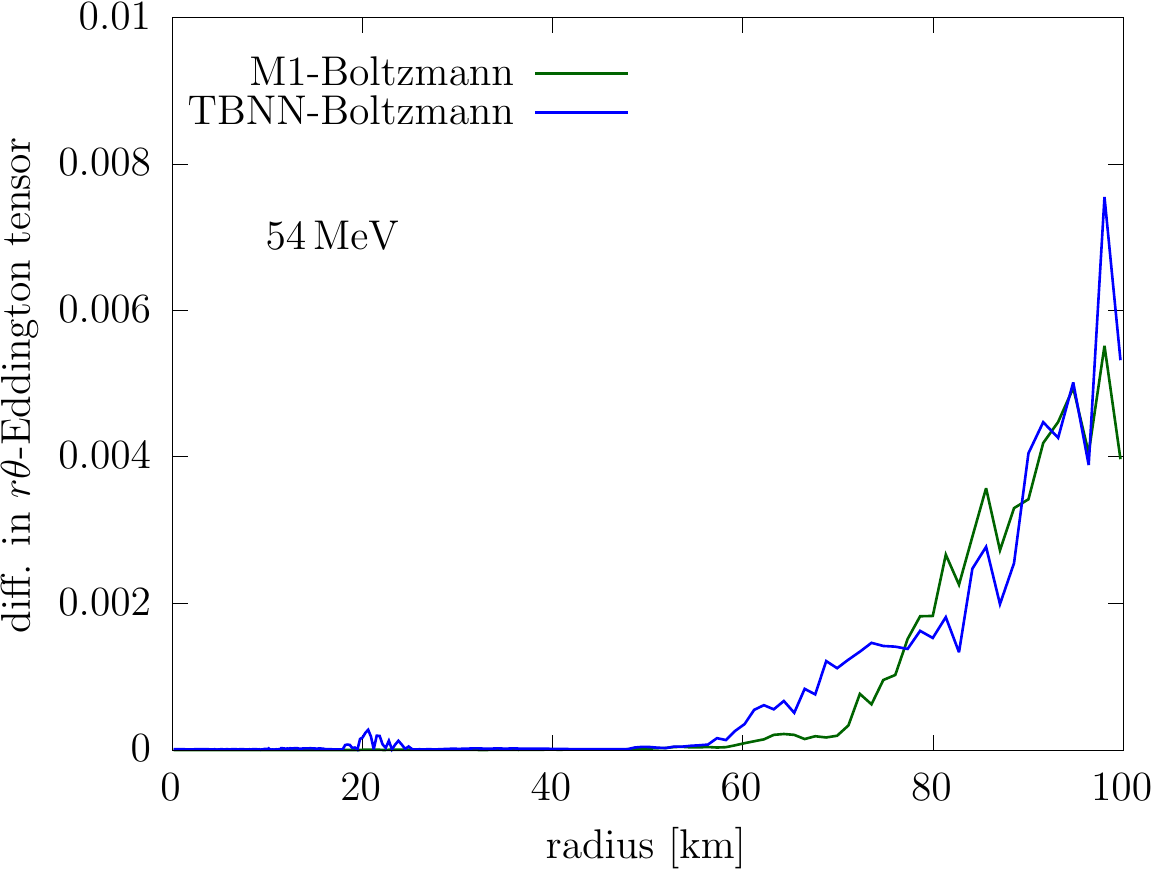} 
\end{tabular}
\caption{The same as figure \ref{fig:CWNNrth_diffline} except that the TBNN-Eddington tensor is shown. \label{fig:TBNNrth_diffline}}
\end{figure*}

Figure \ref{fig:TBNNrth_diffline} shows the TBNN counterpart of figure \ref{fig:CWNNrth_diffline}. The rms difference of the $r\theta$-component from the Boltzmann-Eddington tensor is smaller for the TBNN-Eddington tensor than for the CWNN-Eddington tensor. For the low-energy ($\epsilon=1\,{\rm MeV}$) neutrinos, the difference is typically $\sim 0.002$ for the TBNN, while it is $\sim 0.004$ for the CWNN. In the region $r \la 80\,{\rm km}$, the difference is similar to that of the M1-Eddington tensor, whereas it is smaller at larger radii. This is true also for the intermediate-energy ($\epsilon=8.5\,{\rm MeV}$) neutrinos. For the high-energy ($\epsilon=54\,{\rm MeV}$) neutrinos, the result is not improved from the CWNN. These results indicate that, as a whole, the TBNN reproduces the $r\theta$-component of the Boltzmann-Eddington tensor better than the M1- and CWNN-Eddington tensors.

\subsection{Comparison with other closure relations}
Though the M1-closure relation is popular, there are other closure relations for the moment scheme. Frequently used in CCSN simulations \citep[e.g.,][]{2015MNRAS.453.3386J} are Minerbo \citep{1978JQSRT..20..541M}, Janka \citep{1991PhDT........73J}, and Maximum Entropy Fermi--Dirac \citep[MEFD,][]{1994ApJ...433..250C} closures. In order to see the performance of the TBNN closure relation, we compare it with these closure relations. We employ the TBNN alone here because it performs better than the CWNN as we have just demonstrated.

The closure relations considered in this section are all based on equations (\ref{eq:PM1}, \ref{eq:alpha}, \ref{eq:beta}, \ref{eq:PM1LBthick}, \ref{eq:PM1LBthin}), but their Eddington factors are different from equation (\ref{eq:chi}). The Eddington factors for these closures are given as follows:
\begin{eqnarray}
\chi_{\rm Minerbo} &=& \frac{1}{3}+\frac{1}{15}(6 \tilde{f}^2 - 2\tilde{f}^3 + 6\tilde{f}^4), \\
\chi_{\rm MEFD} &=&\frac{1}{3} + \frac{2}{3}(1-e)(1-2e)\sigma\left(\frac{\tilde{f}}{1-e}\right), \\
\chi_{\rm Janka} &=& \frac{1}{3}\left(1 + \frac{1}{2}\tilde{f}^{1.31} + \frac{3}{2}\tilde{f}^{3.56} \right),
\end{eqnarray}
where the subscripts denote the names of the closure relations. The Minerbo-closure relation is based on the maximum-entropy packing of fermions in momentum space for a given flux factor under an assumption of the vanishing occupation number, which is a dimensionless quantity defined as $e := E/\epsilon^3$, with $E$ being the spectral energy density. The MEFD closure relation extends the Minerbo closure relation to non-zero occupation numbers, which is why it includes $e$ in its expression of the Eddington factor. The function $\sigma$ in the same formula is defined as $\sigma(x) := x^2(3-x+3x^2)/5$. On the other hand, the Janka closure relation is constructed by fitting the results of Monte-Carlo simulations of PNS cooling.

\begin{figure}[t]
\plotone{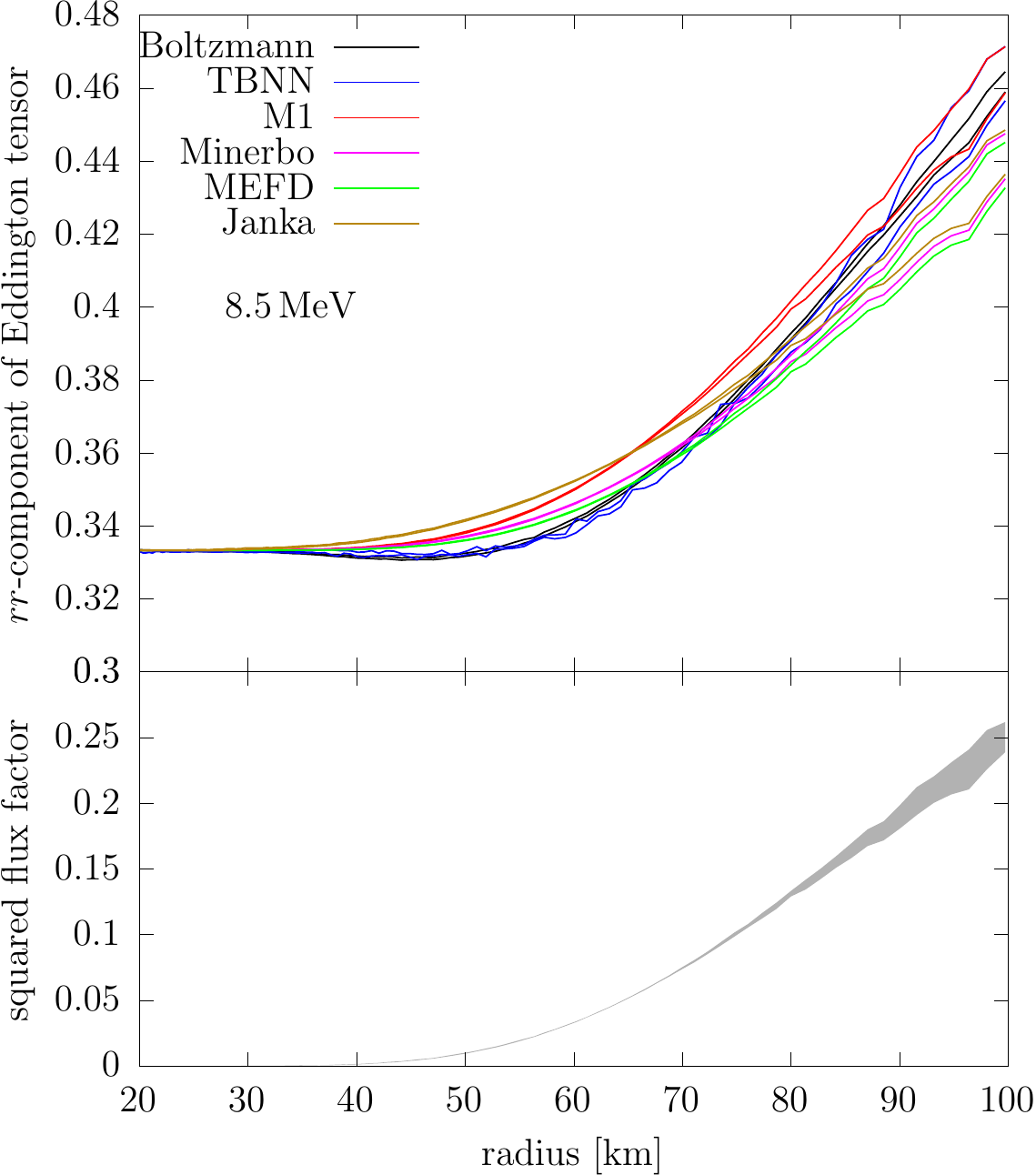}
\caption{Comparison among the different closures and the squared flux factor. The neutrino energy is $8.5\,{\rm MeV}$, and the postbounce time is $100\,{\rm ms}$. (Upper panel) The radial profiles of the $rr$-components of the Eddington tensors given by the Boltzmann simulation (supervisor, black), TBNN (blue), and analytic closures are shown; the red, magenta, green, and yellow represent M1, Minerbo, MEFD, and Janka, respectively. For each closure, two lines indicate the standard deviations from the mean values, i.e., the upper and lower edges of the shaded area in figure \ref{fig:TBNN_fill}. (Lower panel) The radial profile of the squared flux factor $\tilde{f}^2$ is indicated. The shaded band shows the range of the mean plus/minus the standard deviation. \label{fig:closurefluxcomp}}
\end{figure}

The upper panel of figure \ref{fig:closurefluxcomp} shows the comparison among different closure relations. To avoid too much information, we show only the standard deviations from the mean values as a function of radius. The TBNN results agree best with the Boltzmann-Eddington tensor. Among the analytic closure relations, the MEFD closure relation performs better than the M1, Minerbo, and Janka closure relations at small radii. In contrast, the M1 closure relation traces the Boltzmann result better than the others at large radii. Note that the Eddington factor calculated from the distribution function is always larger than the squared flux factor $\tilde{f}^2$, which is based on the simple fact that the distribution function is non-negative \citep[see, e.g., equation (15) of][]{1984JQSRT..31..149L}. The lower panel indicates that all the closure relations considered here satisfy this relation.

Recently, another closure relation based on the Boltzmann simulation was suggested by \cite{2021arXiv210405729N}. They proposed a method to reproduce approximately the angular distribution function of neutrinos from the flux factor. They recognized some correlations between the flux factor and the outgoing/incoming parts of the distribution function in the CCSN simulations with the Boltzmann transport. Using this correlation, they fit a piece-wise parabolic angular distribution to the Boltzmann distribution function. Although their primal interest is the application to collective neutrino oscillations, or more precisely, to the search for the so-called electron neutrino lepton number crossing, which triggers the fast pairwise collective neutrino oscillation, they also provided the Eddington factor based on their reconstructed angular distributions. Their closure relation is close to the Minerbo closure relation at small values of flux factor, while it is more like the M1 closure relation at large values. It is hence expected that their closure relation performs in a similar way to the Minerbo and M1 closure relations in each regime. We stress again that the TBNN closure performs better than both of them. We are afraid that their method is not very accurate in reproducing the incoming part of the distribution function, which also contributes to the Eddington tensor.

\subsection{Generalization to a different time snapshot}
\label{sec:timegen}
The results presented so far are based on a single time snapshot, and hence it is legitimate to ask how well our networks can be applied to other times and other models. In other words, we need to check the generalization performance of our networks. Since this paper is meant for proof of principle, it is much beyond its scope to conduct a full-scale test for many snapshots at different times of different models. Instead, we will be content with the minimum here: we check the performance of the TBNN for another data set at $150\,{\rm ms}$ after core bounce of the same model. This data set is essentially identical to the one described in section \ref{sec:method} except for the snapshot time. Note, however, that unlike the $100\,{\rm ms}$ data, which is split into the training and validation sets, it is entirely the validation set: we input the whole data into the TBNN trained by the $100\,{\rm ms}$ training data. We compare the results of the TBNN with the Boltzmann- and M1-Eddington counterparts. We do not consider the CWNN because the TBNN is demonstrated to be slightly better in section \ref{sec:tbnn}.

\begin{figure*}[t]
\centering
\begin{tabular}{ccc}
\includegraphics[width=0.3\hsize]{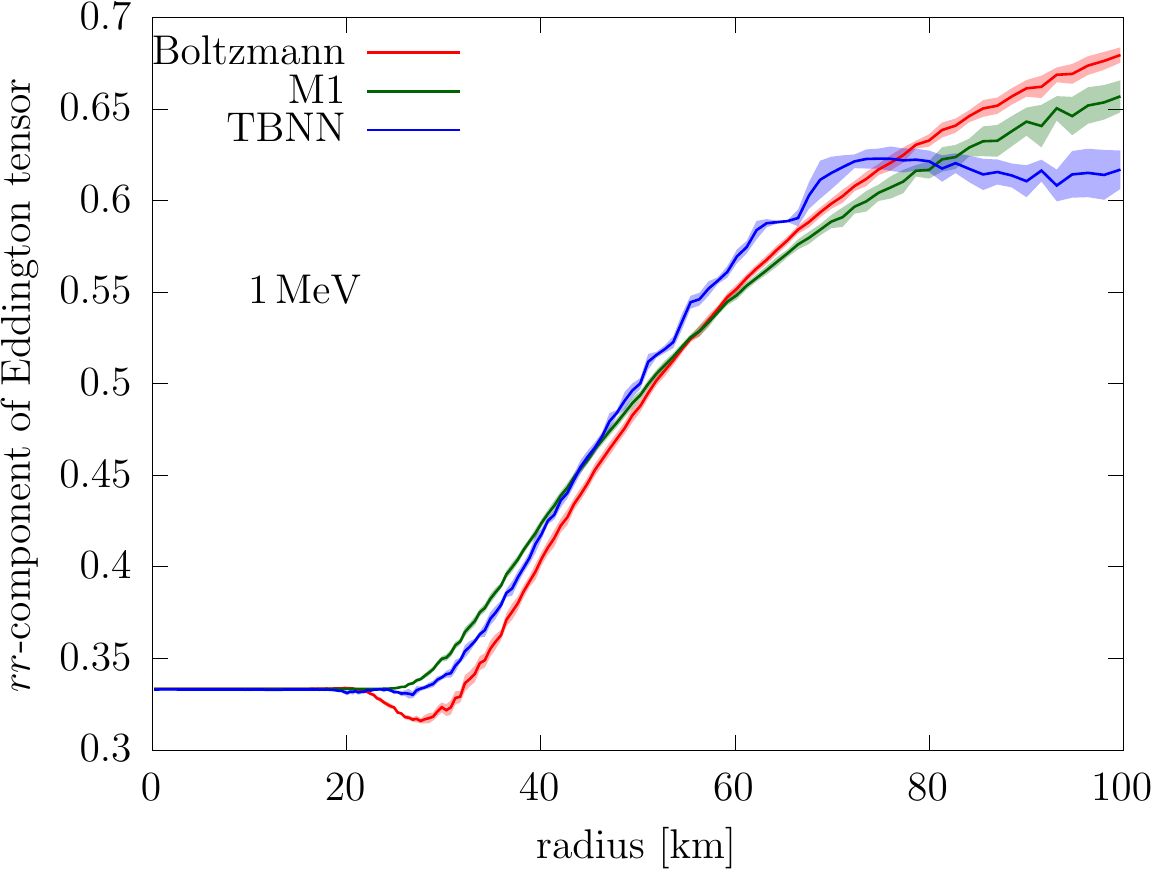} & \includegraphics[width=0.3\hsize]{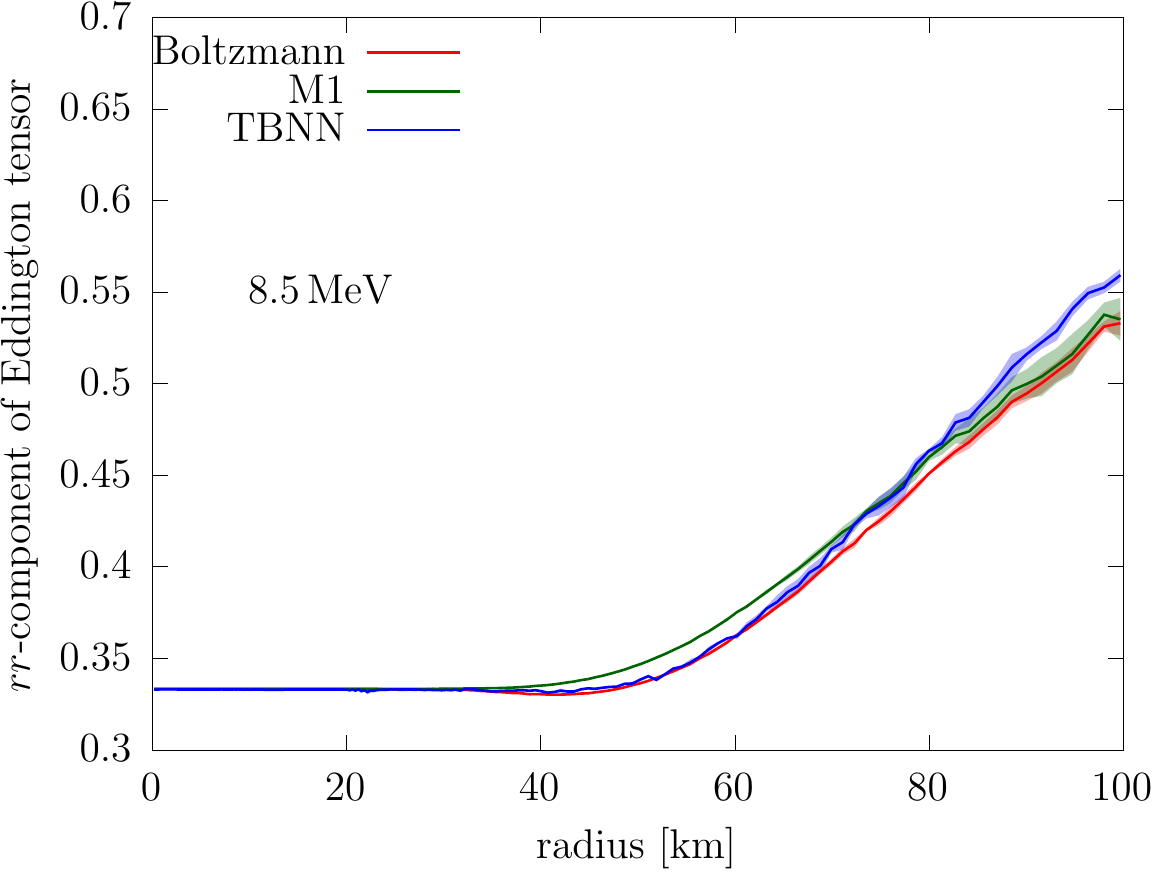} & \includegraphics[width=0.3\hsize]{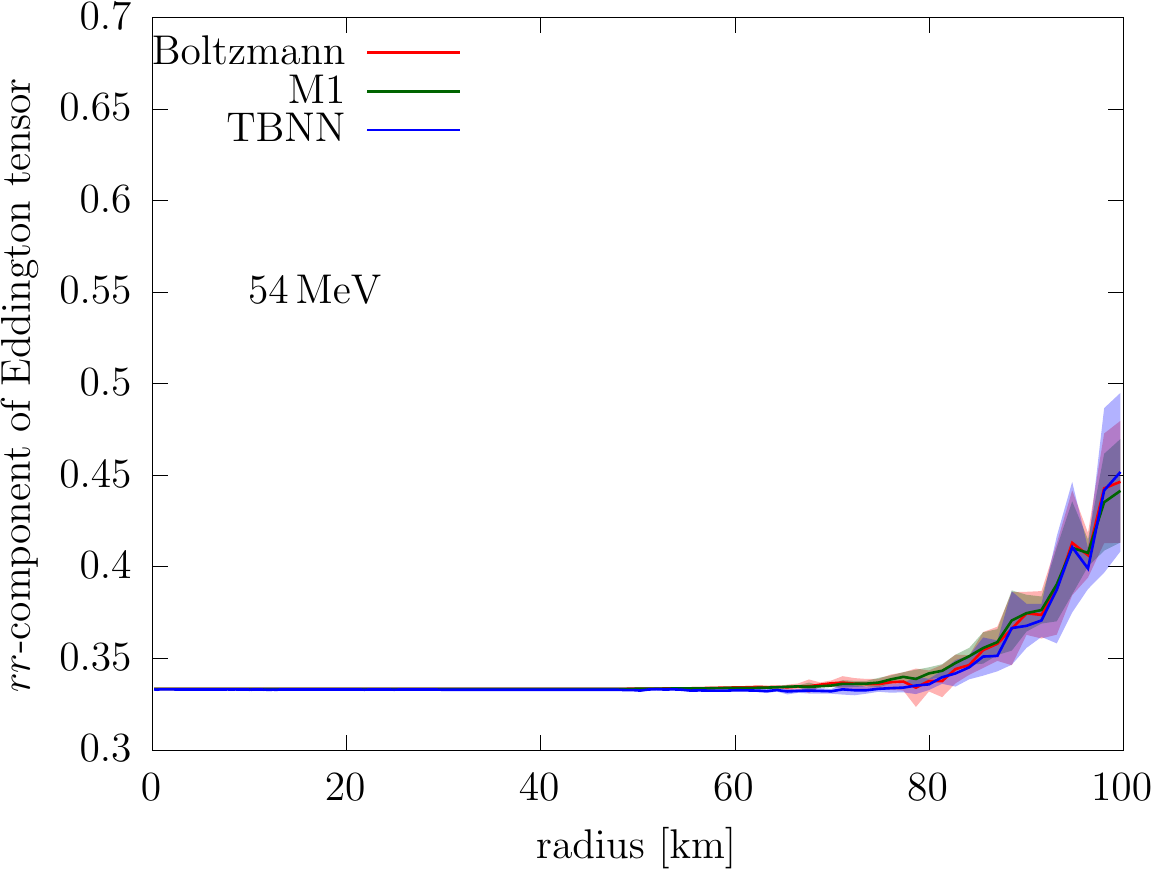} \\
\includegraphics[width=0.3\hsize]{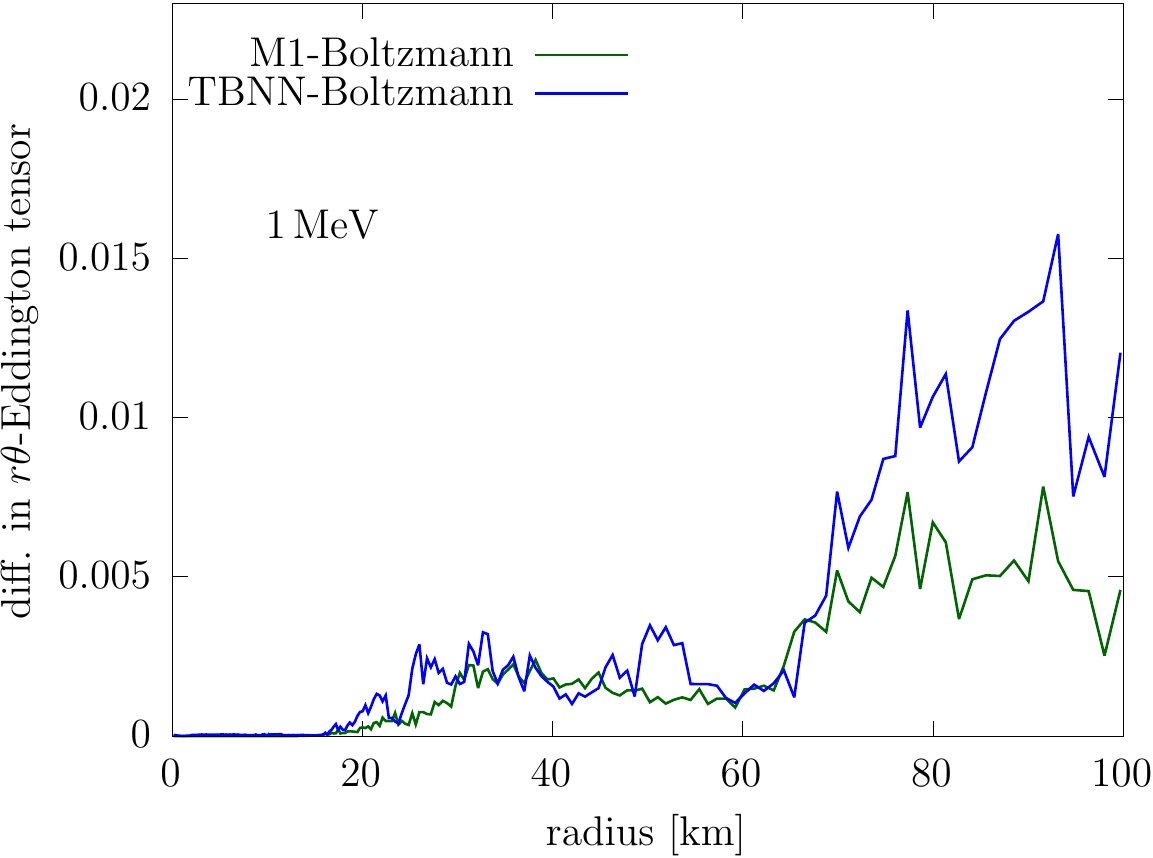} & \includegraphics[width=0.3\hsize]{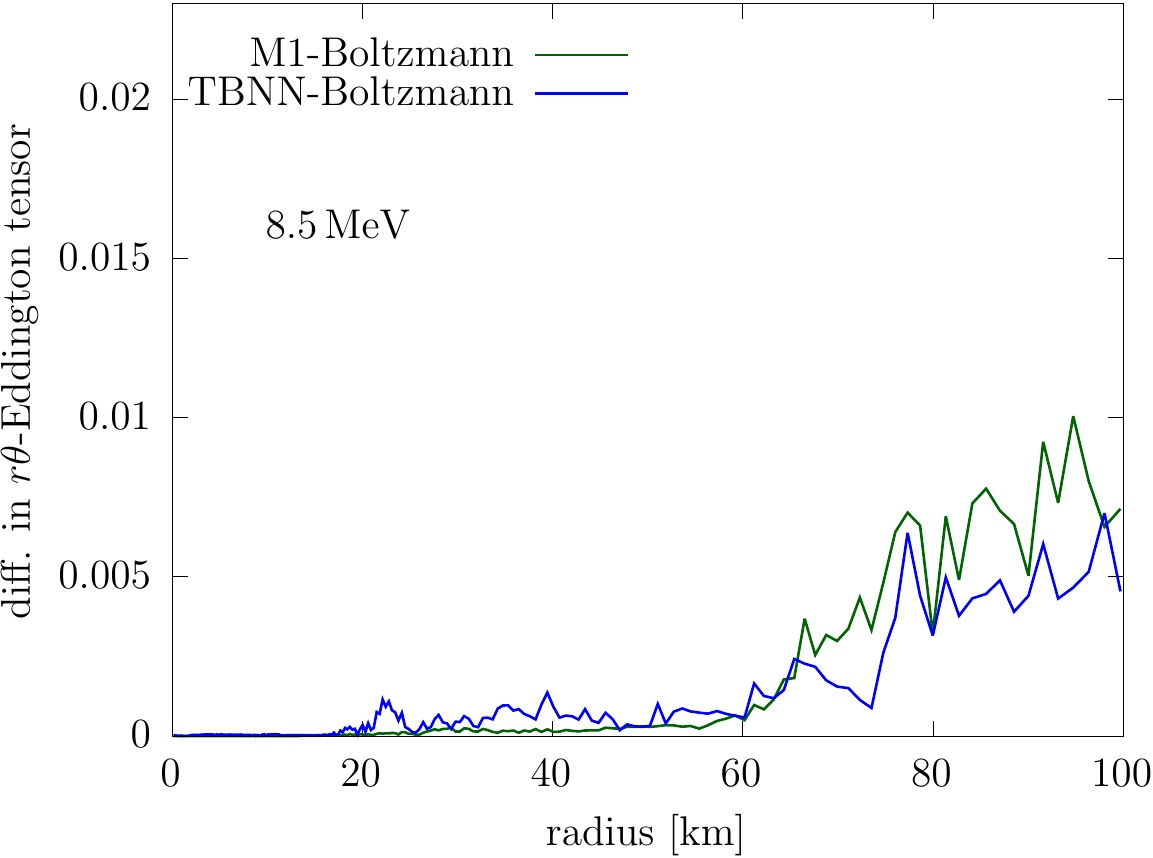} & \includegraphics[width=0.3\hsize]{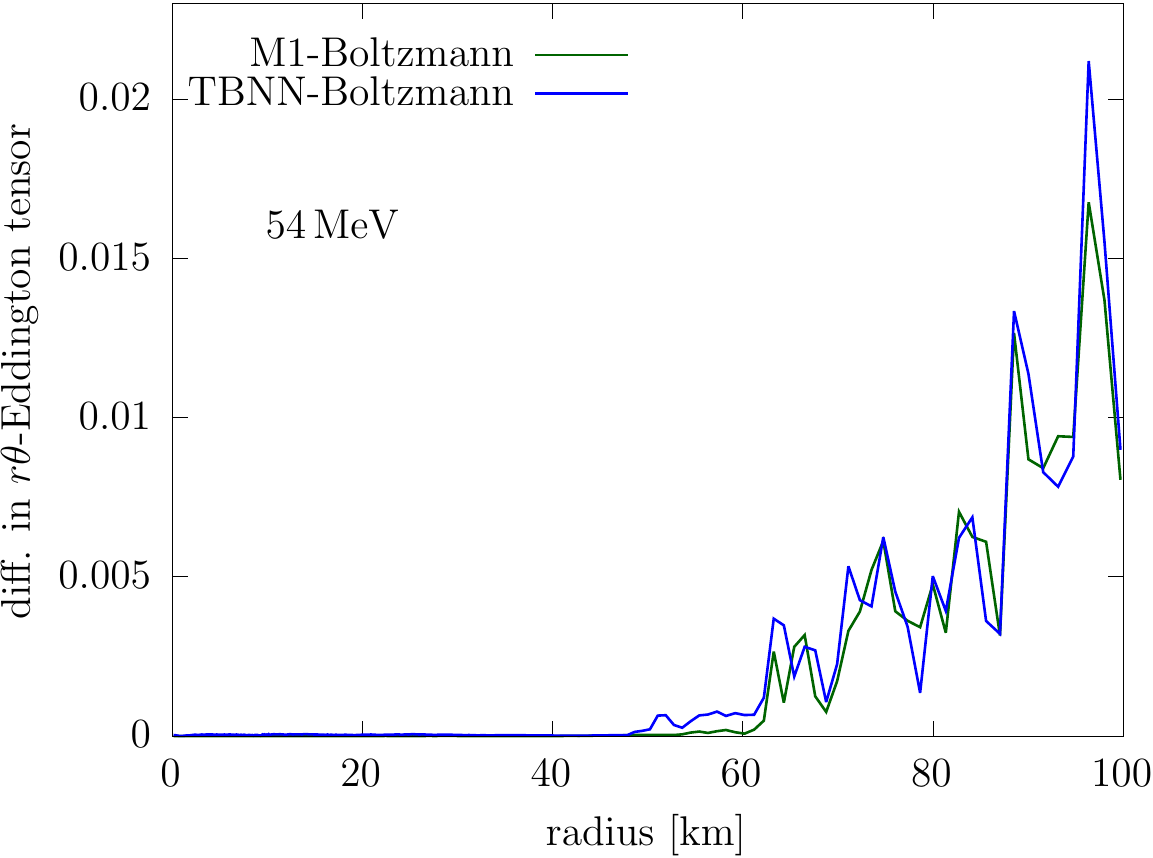} 
\end{tabular}
\caption{\label{fig:gentime} Generalization performance test using the $150\,{\rm ms}$ snapshot. The lines and colors in upper ($rr$-components) and lower (difference of $r\theta$-components from the Boltzmann-Eddington tensor) rows are the same as figures \ref{fig:TBNN_fill} and \ref{fig:TBNNrth_diffline}, respectively.}
\end{figure*}

Figure \ref{fig:gentime} shows the comparison for the $150\,{\rm ms}$ data. It is presented in a similar way to figures \ref{fig:TBNN_fill} and \ref{fig:TBNNrth_diffline}. For the $rr$-components, the TBNN-Eddington tensor is closer to the Boltzmann-Eddington tensor than the M1-Eddington tensor in the inner regions, while the M1-Eddington tensor performs better in the outer regions. In particular, the TBNN-Eddington tensor fails to reproduce the $rr$-components larger than $\sim0.6$. This is just as expected, though, and simply reflects the fact that the sample in the $100\,{\rm ms}$ data is limited to $k^{rr} \la 0.65$ as seen in figure \ref{fig:TBNN_fill}. For the $r\theta$-components, the TBNN performance is similar to or slightly worse than the M1 performance. These results indicate that the training at a single time is not sufficient and we need to expand the training set to include multiple time snapshots. The generalization performance should be checked then for other models with different progenitors with different input physics. That will be the task in the future publication.

\section{Summary and Conclusion} \label{sec:concl}
In order to estimate the Eddington tensor from the lower moments and the fluid velocity, we conducted machine learning. This is the first-ever attempt of the kind and should be regarded as proof of principle that the machine learning closure relation can be constructed from numerical data obtained in the CCSN simulation. We employed two kinds of network structures, the component-wise neural network (CWNN) and the tensor-basis neural network (TBNN). The CWNN is the network that outputs each component of the Eddington tensor individually, while the TBNN calculates the coefficients of the tensors, which are provided to the network as inputs, to output the Eddington tensor as a sum.

We trained these networks with data taken from one of our supernova simulations with the Boltzmann-neutrino transport. In the simulation, the Furusawa-Togashi nuclear equation of state was employed; the nonrotating $15\,M_\odot$ progenitor was adopted; the snapshot at $100\,{\rm ms}$ after the core bounce was chosen. The details of the setup and resultant dynamics will be presented elsewhere. The Eddington tensor and the energy flux and density are calculated directly from the distribution functions for the electron-type neutrinos obtained by the simulation and employed as the output supervisor and the input, respectively, in the training of the networks.

After training the networks successfully, we validated our networks by comparing the output with the supervisor data and the M1-Eddington tensor. They reproduce the dominant component of the Eddington tensor, i.e., the $rr$-component, better than the M1-closure relation, especially for the neutrinos with the mean energy at the shock radius approximately. For this diagonal component, the TBNN achieves slightly better performance than the CWNN whereas both the CWNN and TBNN reproduce the $r\theta$-component of the Boltzmann-Eddington tensor better than the M1-Eddington tensor at larger radii $r\ga 80\,{\rm km}$, while the agreement is slightly worse at smaller radii.
It is worrisome that both the CWNN- and TBNN-Eddington tensors fail to follow the Boltzmann-Eddington tensor better than the M1-Eddington tensor at high energies. The contribution to the energy density from these high-energy neutrinos is rather small, though, because they are exponentially suppressed. We hence expect that the influence of the inaccuracy in the Eddington tensor at high energies on neutrino transport is limited. We also compared the TBNN to other closure relations commonly used in the literature and found that it performs better than them.

Finally, we briefly checked the generalization performance. By using the TBNN trained by the data at $100\,{\rm ms}$, we tested whether it reproduces the Eddington tensor at $150\,{\rm ms}$. We found that our network worked in principle. It was also seen, however, that the network failed for high values of the Eddington factor ($k^{rr}\ga 0.65$) simply because the training data lacked such samples. It did not fare better than the M1 closure relation in reproducing the $r\theta$-component.

Although our ultimate goal is to develop the neutrino-radiation-hydrodynamics code with the moment scheme that implements the DNN closure relation for the simulation of CCSNe, there are many things to do before reaching that stage. We repeat here that this paper is the very first step and its main purpose is proof of principle. Among the issues, we give in the following some brief discussions on the improvement of accuracy and generalization performance of the network, the hydperbolicity of the closure relation the network provides, and the computational cost. They are certainly worth further considerations in the future.

We need to improve the networks in a couple of ways. The expansion of the training data is one of them as discussed in section \ref{sec:timegen}. In fact, the data employed in this paper are taken from a single snapshot of just one simulation for a nonrotating $15\,M_\odot$ progenitor with a baryonic EOS by Furusawa \& Togashi and the standard (but slightly improved) set of the neutrino reactions. Improving our network themselves is another way that we should also try. As a matter of fact, the current networks failed to reproduce some features (e.g., the $rr$-components less than $1/3$ for low neutrino energies). This may be improved, for example, by adding other input data to the current set ($E_{\rm LB}$, $F_{\rm LB}^i$, $V^i$) and/or by adopting other tensor inputs.

In the moment scheme, it is thought to be very important that the closure relation respects the hyperbolicity of the resultant equations. The condition for a moment scheme to be hyperbolic is investigated in \cite{2000MNRAS.317..550P}. For a scheme that gives the Eddington factor $\chi$ only with the flux factor $\tilde{f}$, the sufficient condition is $\chi > \tilde{f}^2$. When the Eddington factor also depends on the energy density, a correction term should be added. Note that $\chi > \tilde{f}^2$ always holds if they are calculated from the same distribution function $f$: it is a consequence of the simple fact that $f>0$. It is hence always satisfied by the flux factor and the Eddington tensor obtained in the Boltzmann simulation. Because the TBNN-Eddington factor $\chi_{\rm TBNN}$ well reproduces the Boltzmann-Eddington factor, it should satisfy $\chi_{\rm TBNN} > \tilde{f}^2$. This can also be checked directly in figure \ref{fig:closurefluxcomp}. As the TBNN Eddington tensor depends not only on the flux but also on the energy density, the hyperbolicity condition may require some corrections. However, from the two facts---it reproduces the Eddington tensor derived from the hyperbolic Boltzmann equation, and the inequality $\chi_{\rm TBNN} > \tilde{f}^2$ holds indeed---the TBNN-Eddington tensor is expected to give a hyperbolic closure relation. More mathematically rigorous investigations, although important, are beyond the scope of this paper.

Finally, the computational cost of the moment scheme radiation transport with the TBNN closure is expected to be much cheaper than the Boltzmann transport. Feedforward in our TBNN, i.e., processing the input data ($E_{\rm LB}$, $F_{\rm LB}^i$, $V^i$) to get the output, $k_{\rm TBNN}^{ij}$, on a single spatial and energy grid takes $\mathcal{O}(10\,{\rm \mu s})$ with GPU (Quadro GV100 by NVIDIA). On the other hand, the Boltzmann solver requires $\mathcal{O}(0.1{\rm s})$ for a single step time evolution per spatial and energy grid on MPI-OpenMP hybrid parallelized CPU (Oakforest-PACS supercomputer). It is true that this is not a fair comparison, but it is obvious that the former is much shorter than the latter. Although this discussion depends strongly on the parallelization settings, the computation of the Eddington tensor should not be a bottleneck of the entire simulation regardless.

\section*{Acknowledgments}

We acknowledge Hideo Matsufuru, Masato Taki, Wakana Iwakami, Enrico Rinaldi, Katsuaki Asano, and Kyohei Kawaguchi for fruitful discussions. This work was supported by Grant-in-Aid for Research Activity Start-up (19K23435) from Japan Society for the Promotion of Science (JSPS), and Grant-in-Aid for Scientific Research on Innovative areas "Gravitational wave physics and astronomy: Genesis" (17H06357, 17H06365) from the Ministry of Education, Culture, Sports, Science and Technology (MEXT), Japan. This work was also supported by MEXT as "Program for Promoting Researches on the Supercomputer Fugaku" (Toward a unified view of the universe: from large scale structures to planets). S. Y. is supported by Institute for Advanced Theoretical and Experimental Physics, Waseda University and the Waseda University Grant for Special Research Projects (project number: 2020-C273). We acknowledge the high-performance computing resources of the K-computer / the supercomputer Fugaku provided by RIKEN, the FX10 provided by Tokyo University, the FX100 provided by Nagoya University, the Grand Chariot provided by Hokkaido University, and Oakforest-PACS provided by JCAHPC through the HPCI System Research Project (Project ID: hp130025, 140211, 150225, 150262, 160071, 160211, 170031, 170230, 170304, 180111, 180179, 180239, 190100, 190160, 200102, 200124) for producing and processing the supervisor data.

%


\software{Tensorflow \citep{tensorflow2015-whitepaper}, keras \citep{chollet2015keras}, gnuplot \citep{gnuplot5}}






\bibliography{ref}{}
\bibliographystyle{aasjournal}



\end{document}